\documentclass[twocolumn,journal]{IEEEtran}
\usepackage{wrapfig}
\usepackage{graphicx}
\usepackage{array}
\usepackage{amsopn}
\usepackage{mathrsfs}		
\usepackage{amssymb,latexsym,amsfonts,amsmath}
\usepackage{subfigure}
\usepackage{calc}
\usepackage{xcolor}
\usepackage{color,soul}
\definecolor{lightblue}{rgb}{.88,1,1}
\sethlcolor{lightblue}
\usepackage{amsthm}
\usepackage{amsopn}
\usepackage{epsfig}			
\usepackage{caption}
\usepackage{algorithm}
\usepackage{color,xcolor,soul}
\usepackage{mathtools,lipsum,cuted}
\usepackage{enumerate}
\usepackage{epstopdf}
\usepackage{stfloats}
\usepackage[unicode=true,
 bookmarks=true,bookmarksnumbered=true,bookmarksopen=true,bookmarksopenlevel=1,
 breaklinks=false,pdfborder={0 0 0},backref=false,colorlinks=false]
 {hyperref}

\usepackage{cuted}
\usepackage{mathtools}
\hypersetup{pdftitle={Your Title},
 pdfauthor={Your Name},
 pdfpagelayout=OneColumn, pdfnewwindow=true, pdfstartview=XYZ, plainpages=false}
\IEEEoverridecommandlockouts
\DeclareRobustCommand*{\lyxarrow}{%
\@ifstar
{\leavevmode\,$\triangleleft$\,\allowbreak}
{\leavevmode\,$\triangleright$\,\allowbreak}}

\theoremstyle{plain}

\theoremstyle{definition}

\theoremstyle{plain}

\newtheorem*{proof:}{Proof}
\newtheorem{theorem}{Theorem}
\newtheorem{lemma}{Lemma}

\newtheorem{definition}{Definition}
\newtheorem{assumption}{Assumption}
\newtheorem{remark}{Remark}

\providecommand{\definitionname}{Definition}
\providecommand{\lemmaname}{Lemma}
\providecommand{\theoremname}{Theorem}

\begin{document}
\title{\textcolor{black}{Analysis of Two-Dimensional Feedback Systems over Networks Using Dissipativity}}
\author{\IEEEauthorblockN{Yang Yan\textsuperscript{*}, Lanlan Su\textsuperscript{*}, Vijay Gupta and Panos Antsaklis}

\thanks{Partial support from the National Science Foundation under Grants No. CNS-1035655, CNS-1446288 and CNS-1544724 and ARO under Grant No. W911NF-17-1-0072 is gratefully acknowledged.

Yan, Su, Gupta and Antsaklis are with the Department of Electrical Engineering, University of Notre Dame, Notre Dame, IN 46556 USA (e-mail: ginger.yang.yan@gmail.com, lanlansu.work@gmail.com, vgupta2@nd.edu, antsaklis.1@nd.edu).
}
\thanks{$*$ These authors contributed equally to this work.}
}
\maketitle

\setstcolor{red}
\begin{abstract}
This paper investigates the closed-loop $\mathcal{L}_2$ stability of two-dimensional (2-D) feedback systems across a digital communication network by introducing the tool of dissipativity. First, sampling of a continuous 2-D system is considered and an analytical characterization of the $QSR$-dissipativity of the sampled system is presented. 
Next, the input-feedforward output-feedback passivity (IF-OFP), a simplified form of $QSR$-dissipativity, is utilized to study the framework of feedback interconnection of two 2-D systems over networks.
Then, the effects of signal quantization in communication links on dissipativity degradation of the 2-D feedback quantized system is analyzed. 
Additionally, an event-triggered mechanism is developed for 2-D networked control systems while maintaining $\mathcal{L}_2$ stability of the closed-loop system. In the end, an illustrative example is provided. 
\end{abstract}


%
\IEEEpeerreviewmaketitle
\IEEEdisplaynontitleabstractindextext

\section{Introduction}
\label{sec:introduction}

Two-dimensional (2-D) continuous systems arise naturally in a wide range of applications, where the system variables depend on two variables, such as time and distance, or height and width. Examples include filter design~\cite{Lu92}, digital image processing \cite{Huang81}, analysis of repetitive processes \cite{Rogers07}, thermal engineering \cite{Benzaouia15}, automobile platoons \cite{Knorn13}, see \cite{Bose17} \cite{Rogers17} for more examples. Accordingly, the study of 2-D systems has received significant attention in the past decades,  starting from the early works of Roesser \cite{Roesser75}, who generalized the state-space realization to 2-D space, and Fornasini \cite{Fornasini78}, who studied stability of 2-D systems. More recent works have considered control of 2-D systems for disturbance attenuation (see e.g., \cite{DU02} \cite{Andrea03} \cite{Scherer216} and the references therein).

It is well-known that the coupling between the vertical state and horizontal state makes the analysis for 2-D systems essentially more difficult than 1-D systems. In this work, we choose a dissipativity based framework to analyze 2-D feedback systems across a digital network.
%
%
Dissipativity (and its special case of passivity) are rather standard notions for 1-D systems~\cite{Willems72}. They provide an energy-based perspective for analysis and design, and relate strongly to Lyapunov and $\mathcal{L}_2$ stability theories \cite{Khalil00}. For 2-D systems, some relevant works are \cite{Ahn15} \cite{Yang08} \cite{Li12} \cite{Li12Robust} that analyzed 2-D systems and designed 2-D filters using dissipativity and passivity. On the other hand, for the massive needs for control over digital networks, for example the feedback control of transport phenomena (momentum, heat and mass transfer) coupled with chemical reactions, the dissipativity-based design allows for simple design of stable complex 2-D systems. 
However, to the best of our knowledge, the existing approaches for 2-D systems are not suitable to analyze the closed-loop stability of networked 2-D systems, which is the main concern of our work. Note that it is not trivial to extend dissipativity-based methods from 1-D systems to 2-D systems. Specifically, since dissipativity and passivity relate a time-dependent input-output supply function to a state-based storage function, dissipativity of a 2-D system is not equivalent to dissipativity of two 1-D systems (corresponding to the horizontal and vertical domains of the 2-D system). Moreover, bridging the dissipativity of a continuous 2-D system and its sampled system, and grouping the ``local error'' in the intermittent communication between a 2-D system and a 2-D controller in an event-triggering framework differ significantly from 1-D systems.

For linear 2-D systems, both the Roesser and F-M model are frequently used, which naturally appear in many practical applications. Since the F-M model can be transformed into Roesser model by model transformation \cite{Benzaouia15}, we focus on the general Roesser model in this work.
The main contribution of this work is to present $\mathcal{L}_2$ stability analysis of 2-D continuous feedback systems across a digital network using dissipativity-based notions. We include three effects of the digital network -- sampling and quantization of sensor output before transmission to the controller and the control input before transmission to the actuator, as also transmissions according to an event triggered scheme. We begin by defining $QSR$-dissipativity for continuous and discrete 2-D systems and identifying the relation of dissipativity to $\mathcal{L}_{2}$-stability. Then, we provide an analytical characterization of the $QSR$-dissipativity of the system that is obtained when the output from the sensor is sampled and transmitted to the controller, and the control input is applied to the system via a sample and zero-order hold mechanism. We show that a large sampling period may lead to the sampled system becoming non-dissipative and for a linear continuous 2-D system, we provide a LMI-based sufficient condition to determine whether the sampled system is $QSR$-dissipative. Next, the input-feedforward output-feedback passivity (IF-OFP), a simplified form of $QSR$-dissipativity, is utilized to study the framework of feedback interconnection of two 2-D systems over network. We include the effect of quantization of the sensor output and the control input before transmission over the network. Following the work on quantization effects in 1-D systems \cite{Garcia13}, we focus on logarithmic quantizers. The passivity degradation caused by the input and the output quantizers is analyzed for the 2-D closed-loop system, and a condition is provided to ensure that the closed-loop system under sampling and quantization remains $\mathcal{L}_2$-stable. Finally, we propose an event-triggered scheme for transmission in a way that reduces the number of transmissions while maintaining $\mathcal{L}_2$-stability of the closed-loop system.

The rest of this paper is organized as follows: Definitions of $QSR$-dissipativity for 2-D systems and the problem formulation are provided in Section \ref{sec:2-D preliminaries}. Section \ref{sec:main results} presents our main results. Section~\ref{sec:2-D Approximation} analyzes dissipassivity of a sampled 2-D system, Section~\ref{sec:quantization} considers the effect of quantization of the sampled sensor output and control input, and Section~\ref{sec:2-D event} addresses the event-triggered transmission of these data. The results are illustrated with the help of an example in Section \ref{sec:2-D example}. Section \ref{sec:2-D conclusion} concludes this work. 

\textbf{Notation:} The sets of real numbers, non-negative real numbers, and non-negative integers are denoted by $\mathbb{R}$, $\mathbb{R}^{+}$ and $\mathbb{N}$ respectively. The Euclidean space of dimension $n$ is denoted by $\mathbb{R}^n$. The Euclidean norm is denoted by $\left|\cdot\right|$. Denote $\mathcal{L}_{2}^{n}$ as the set of signals $f:[0,\infty)\times[0,\infty)\rightarrow\mathbb{R}^{n}$ satisfying $\mathcal{L}_{2}^{n}=\left\{ f(z_{1},z_{2}):||f(z_{1},z_{2})||_{\mathcal{L}_{2}}^{2}=\int_{0}^{\infty}\int_{0}^{\infty}|f(z_{1},z_{2})|^{2}dz_{1}dz_{2}<\infty\right\}$ when $f$ is continuous on its domain and $\mathcal{L}_{2}^{n}=\left\{ f(i,j):||f(i,j)||_{\mathcal{L}_{2}}^{2}=\sum_{i=0}^{\infty}\sum_{j=0}^{\infty}|f(i,j)|^{2}<\infty\right\}$ when $f$ is discrete on its domain. For a symmetric matrix $A$, the minimum eigenvalue of $A$ is denoted as $\underline{\lambda}(A)$ and the maximum eigenvalue is denoted as $\bar{\lambda}(A)$. The $n$-dimensional identity matrix is denoted by $I_{n\times n}$, or simply by $I$ if its dimension is clear from the context. The notation $\text{max}(a,b)$ represents the larger of the values between $a,b\in\mathbb{R}$. The Kronecker product of two matrices $A$ and $B$ is denoted by $A\otimes B$. For a symmetric matrix represented blockwise, off diagonal blocks are abbreviated with ``*''.

\section{Backgrounds}
\label{sec:2-D preliminaries}
\subsection{Preliminaries}

Consider a nonlinear continuous 2-D system $G_p$ in the following form:
\begin{equation}
\left\{ \begin{array}{c}
\frac{\partial x_{h}(z_1,z_2)}{\partial z_1}=f_{h}(x_{h}(z_1,z_2),x_{v}(z_1,z_2),u(z_1,z_2))\\
\frac{\partial x_{v}(z_1,z_2)}{\partial z_2}=f_{v}(x_{h}(z_1,z_2),x_{v}(z_1,z_2),u(z_1,z_2))\\
y(z_1,z_2)=g(x_{h}(z_1,z_2),x_{v}(z_1,z_2),u(z_1,z_2))
\end{array}\right.
\label{eq:2-D nonlinear 2-D system}
,\end{equation}
where $z_1,z_2\ge0$ are two independent variables. The variable $z_1$ is termed as the horizontal coordinate, while $z_2$ is termed as the vertical coordinate. In the above equation defined through the state space $\mathcal{X}_h$, $\mathcal{X}_v$, the input space $\mathcal{U}$ and the output space $\mathcal{Y}$, $x_{h}\in\mathcal{X}_{h}\subset\mathbb{R}^{n_{h}}$ is called the horizontal state of the system, $x_{v}\in\mathcal{X}_{v}\subset\mathbb{R}^{n_{v}}$ is called the vertical state of the system. The variable $u\in\mathcal{U}\subset\mathbb{R}^{m}$ denotes the input to the system while  $y\in\mathcal{Y}\subset\mathbb{R}^{p}$ is the output of the system. 
The boundary condition for the system (\ref{eq:2-D nonlinear 2-D system}) is specified by the pair $x_{h}(0,z_2),x_{v}(z_1,0)$. 

A special case of~(\ref{eq:2-D nonlinear 2-D system}) is when the system is linear. We will sometimes focus on this case that is called the 2-D Roesser model as described by 
\begin{equation}
\left\{ \begin{array}{c}
\frac{\partial x_{h}(z_1,z_2)}{\partial z_1}=A_{11}x_{h}(z_1,z_2)+A_{12}x_{v}(z_1,z_2)+B_{1}u(z_1,z_2)\\
\frac{\partial x_{v}(z_1,z_2)}{\partial z_2}=A_{21}x_{h}(z_1,z_2)+A_{22}x_{v}(z_1,z_2)+B_{2}u(z_1,z_2)\\
y(z_1,z_2)=C_{1}x_{h}(z_1,z_2)+C_{2}x_{v}(z_1,z_2)+Du(z_1,z_2).
\end{array}\right.\label{eq:2-DCTsystem}
\end{equation}

Given the coupling between the horizontal and vertical states, we cannot define $QSR$-dissipativity of the system (\ref{eq:2-D nonlinear 2-D system}) through dissipativity of each state separately. Instead, we propose the following definition.
\begin{definition}
\label{def:2-D $QSR$} 
Given real matrices $Q,S,R$ of appropriate size, the continuous 2-D system \eqref{eq:2-D nonlinear 2-D system} is said to be \emph{QSR-dissipative} if there exists a non-negative function $V=V_h+V_v$ where $V_h:\mathcal{X}_{h}\rightarrow\mathbb{R}^{+}$ and $V_v:\mathcal{X}_{v}\rightarrow\mathbb{R}^{+}$, called \textup{storage function}, such that for all $Z_1\ge0,Z_2\ge0$, all $x_{h}(0,z_2)\in\mathcal{X}_{h}$, $x_{v}(z_1,0)\in\mathcal{X}_{v}$, and all input functions $u\in\mathcal{U}$, the following inequality holds
\begin{align}
\label{eq:2-D $QSR$}
\begin{split}
& \quad \int_0^{Z_1} \lbrack V_v(x_v(z_1,Z_2)) - V_v(x_v(z_1,0)) \rbrack dz_1 \\
& \quad +\int_0^{Z_2} \lbrack V_h(x_h(Z_1,z_2)) - V_h(x_h(0,z_2)) \rbrack dz_2 \\ 
&\leq 
\int_{0}^{Z_1}\int_{0}^{Z_2}\text{ }[y^{T}(\text{\ensuremath{\tau}}_{1},\tau_{2})Qy(\text{\ensuremath{\tau}}_{1},\tau_{2})+2y^{T}(\text{\ensuremath{\tau}}_{1},\tau_{2})Su(\text{\ensuremath{\tau}}_{1},\tau_{2})\\
&\quad +u^{T}(\text{\ensuremath{\tau}}_{1},\tau_{2})Ru(\text{\ensuremath{\tau}}_{1},\tau_{2})]d\tau_{1}d\tau_{2},
\end{split}
\end{align}
along the trajectories of the system.
\end{definition}
Definition~\ref{def:2-D $QSR$} extends the standard dissipativity inequality from 1-D systems to 2-D systems and has a similar intuitive explanation that given any $Z_1,Z_2\ge0,$ the energy stored in the final boundary state, $x_{h}(Z_1,z_2)$, $z_2\in[0,Z_2]$ and $x_{v}(z_1,Z_2)$, $z_1\in[0,Z_1]$, is at most equal to the sum of the energy stored in the initial boundary condition $x_{h}(0,z_2)$ for $z_2\in[0,Z_2]$, $x_{v}(z_1,0)$ for $z_1\in[0,Z_1]$, and the total external supplied energy. Compared with the definition of $QSR$-$\alpha$-dissipativity defined in \cite{Ahn15} for 2-D systems with zero initial condition, Definition \ref{def:2-D $QSR$} in this paper adopts a storage function which examines the changing amount of stored energy of the dynamical 2-D system. Using a storage function, it enables us to consider the case with non-zero boundary conditions. It should be noted that the form of the storage function in the left hand side of \eqref{eq:2-D $QSR$} with two coordinates separated is relatively conservative. It is true that less or non-conservative Lyapunov functions have been proposed for stability analysis for 2-D systems {\cite{Willems07}} {\cite{Chesi14}}. However, restricting its form in this way makes the dissipativity analysis of 2-D systems easier. This will be illustrated by Theorem 2 in Section III-A.

Given that we are interested in discrete 2-D systems that are obtained by sampling a continuous 2-D system, we also define dissipativity of a nonlinear discrete 2-D system described as
\begin{equation}
\left\{ \begin{array}{c}
x_{h}(i+1,j)=f_{h}(x_{h}(i,j),x_{v}(i,j),u(i,j))\\
x_{v}(i,j+1)=f_{v}(x_{h}(i,j),x_{v}(i,j),u(i,j))\\
y(i,j)=g(x_{h}(i,j),x_{v}(i,j),u(i,j))
\end{array}\right.
\label{eq:2-D discrete nonlinear system}
,\end{equation}
where $x^h(i,j) \in \mathcal{X}_{h} \in\mathbb{R}^{n_h}$ is the horizontal state, $x_v(i,j)\in \mathcal{X}_{v} \in \mathbb{R}^{n_v}$ is the vertical state, $u(i,j) \in \mathcal{U}\in\mathbb{R}^m$ is the input and $y(i,j)\in \mathcal{Y}\in\mathbb{R}^{p}$ is the output.

\begin{definition}\label{def:2-D DT}
Given real matrices $Q,S,R$ of appropriate size, the discrete 2-D system (\ref{eq:2-D discrete nonlinear system}) is said to be \emph{QSR-dissipative} if there exists a non-negative storage function  $V=V_h+V_v$ where $V_h:\mathcal{X}_{h}\rightarrow\mathbb{R}^{+}$ and $V_v:\mathcal{X}_{v}\rightarrow\mathbb{R}^{+}$,  such that for all $N_1,N_2\in \mathbb{N}$, all $x_{h}(0,j)\in\mathcal{X}_{h}$, $x_{v}(i,0)\in\mathcal{X}_{v}$, and all input functions $u\in\mathcal{U}$,
\begin{align}
\begin{split}
    & \quad \sum_{i=0}^{N_1-1} \lbrack V_v(x_v(i,N_2)) - V_v(x_v(i,0)) \rbrack \\
    & \quad +\sum_{j=0}^{N_2-1} \lbrack V_h(x_h(N_1,j))- 
     V_h(x_h(0,j)) \rbrack \\
    &\leq
    \sum_{i=0}^{N_1-1}\sum_{j=0}^{N_2-1}\text{ }[y^{T}(i,j)Qy(i,j)+2y^{T}(i,j)Su(i,j)\\
    &\quad  +u^{T}(i,j)Ru(i,j)],
\label{eq:2-D $QSR$ disrete}
\end{split}
\end{align}
along the trajectories of the system.
\end{definition}

Since $y^{T}Qy=\frac{1}{2}y^T(Q+Q^{T})y$ and $u^{T}Ru=\frac{1}{2}u^{T}(R+R^{T})u$, we restrict the matrices $Q$ and $R$ in Definition \ref{def:2-D $QSR$} and \ref{def:2-D DT} to be symmetric  without loss of generality.  

By generalizing the definition of $\mathcal{L}_2$ stability for 1-D systems {\cite{van00}}, a continuous 2-D system is $\mathcal{L}_2$ stable, if there is a constant gain $\gamma>0$ and $\beta$, such that for all $Z_1,Z_2\in\mathbb{R}^+$, all $x_h(0,z_2)\in\mathcal{X}_h$, $x_v(z_1,0)\in\mathcal{X}_v$, and all $u\in\mathcal{U}$,
\begin{equation*}
\begin{array}{l}
    \label{eq:L2 stable continuous}
    \int_{0}^{Z_1} \int_{0}^{Z_2} \left|y_(z_1,z_2)\right|^2 dz_2dz_1 \\
    \leq \gamma^2   \int_{0}^{Z_1} \int_{0}^{Z_2} \left|u(z_1,z_2) \right|^2 dz_2dz_1 + \beta.
\end{array}\end{equation*}
Consider the $QSR$-dissipative state-space system (\ref{eq:2-D nonlinear 2-D system}) with $Q=-I$, $S=0$, $R=\gamma^2I$, we can observe that it is $\mathcal{L}_2$ stable with $\beta = \int_{0}^{Z_1} V_v(x_v(z_1,0))dz_1+\int^{Z_2}_0 V_h(x_h(0,z_2))dz_2$. Similar statement can be made for the discrete case.

Passivity and some related definitions can now be given as follows. 
\begin{definition}
\label{def:passivity}
Consider a continuous 2-D system \eqref{eq:2-D nonlinear 2-D system} or a discrete 2-D system \eqref{eq:2-D discrete nonlinear system} that is $QSR$-dissipative. 
The system 
\begin{itemize}
\item is \emph{passive}, if the system is $(0,\frac{1}{2}I,0)$-dissipative; $ $
\item is \emph{input feedforward passive (IFP)}, if there exists a constant $\nu\in\mathbb{R}$ such that the system is $(0,\frac{1}{2}I,-\nu I)$-dissipative; we call such a $\nu$ an IFP level, denoted as IFP$(\nu)$; 
\item is \emph{output feedback passive (OFP)}, if there exists a constant $\rho\in\mathbb{R}$ such that the system is $(-\rho I,\frac{1}{2}I,0)$-dissipative; we call such a $\rho$ an OFP level, denoted as OFP$(\rho)$;

\item is \emph{input-feedforward-output-feedback passive (IF-OFP)}, if the system is $(-\rho I,\frac{1}{2}I,-\nu I)$-dissipative for some $\rho,\nu\in\mathbb{R}$; we call such $\rho$ and $\nu$   IF-OFP levels, denoted as IF-OFP$(\rho,\nu)$; 
\item has a \emph{finite $\mathcal{L}_{2}$-gain}, if there exists a constant
$\gamma > 0 $ such that the system is $(-I,0,\gamma^2 I)$-dissipative,
denoted as FGS$(\gamma)$.
\end{itemize}
\end{definition}
\begin{figure}
\centering
 \captionsetup{font={scriptsize}}
 \includegraphics[scale = 0.61]{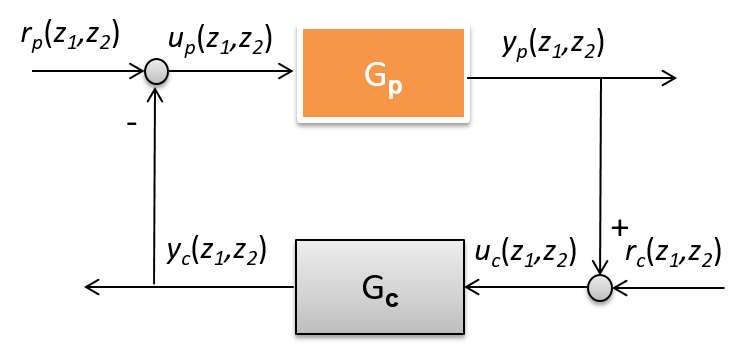}
 \caption{Feedback Interconnection of Two 2-D Systems}
 \label{fig:feedback interconnection}
\end{figure}

For both continuous and discrete 2-D systems (\ref{eq:2-D nonlinear 2-D system}) and (\ref{eq:2-D discrete nonlinear system}), the following results state the relationship between $QSR$-dissipativity and $\mathcal{L}_2$ stability. The proofs of these results follow along those of similar results for 1-D systems \cite{Khalil00} \cite{Hill76} and are omitted.
\begin{lemma}
\label{lem:2-D Q<0 implies l2 stable}
A system that is $QSR$-dissipative for $Q<0$ is $\mathcal{L}_2$-stable.
\end{lemma}

\begin{lemma}
\label{lem:2-D closed loop stable using index}
Consider the feedback interconnection depicted in  Figure \ref{fig:feedback interconnection} where each of the 2-D feedback subsystems $G_p$ and $G_c$ are IF-OFP$(\nu_1,\rho_1)$ and IF-OFP$(\nu_2,\rho_2)$, respectively. Then the closed-loop system is $\mathcal{L}_2$-stable if
\begin{equation}
    \begin{split}
        \label{eq:v+p>0}
        \nu_2+\rho_1>0, \\
        \nu_1+\rho_2>0.
    \end{split}
\end{equation}
\end{lemma}

\subsection{Problem Formulation}
\label{sec:2-D problem formulation}
As shown in Figure \ref{fig:feedback interconnection}, let us consider the feedback interconnection of 2-D continuous systems $G_{p}$ and $G_{c}$ across a digital communication network. 
In particular, let the output of the system $G_{p}$ be sampled and quantized before transmission across the network according to an event triggered mechanism. Similarly, let the control input from $G_{c}$ be quantized before transmission across the network and applied back to the system $G_{p}$ through a zero order hold. It is known that all these operations -- sampling, quantization, and event-triggered transmission -- can lead to loss of dissipativity.
In this paper, we wish to identify conditions under which 2-D systems retain their dissipativity properties under these operations, and further, the impact of these operations on the $\mathcal{L}_{2}$ stability of the closed-loop system.  

Particularly, the problem can be divided into three steps. The first step is to characterize the $QSR$-dissipativity of a sampled 2-D system. Then by utilizing IF-OFP, a simplified form of $QSR$-dissipativity, the second step is to analyze the effect of the quantization on the closed-loop $\mathcal{L}_2$ stability. The third step is to design an event-triggered mechanism for 2-D networked control systems while maintaining $\mathcal{L}_2$ stability of the closed-loop system. 


\begin{figure}
\centering
 \captionsetup{font={scriptsize}}
 \includegraphics[scale = 0.51]{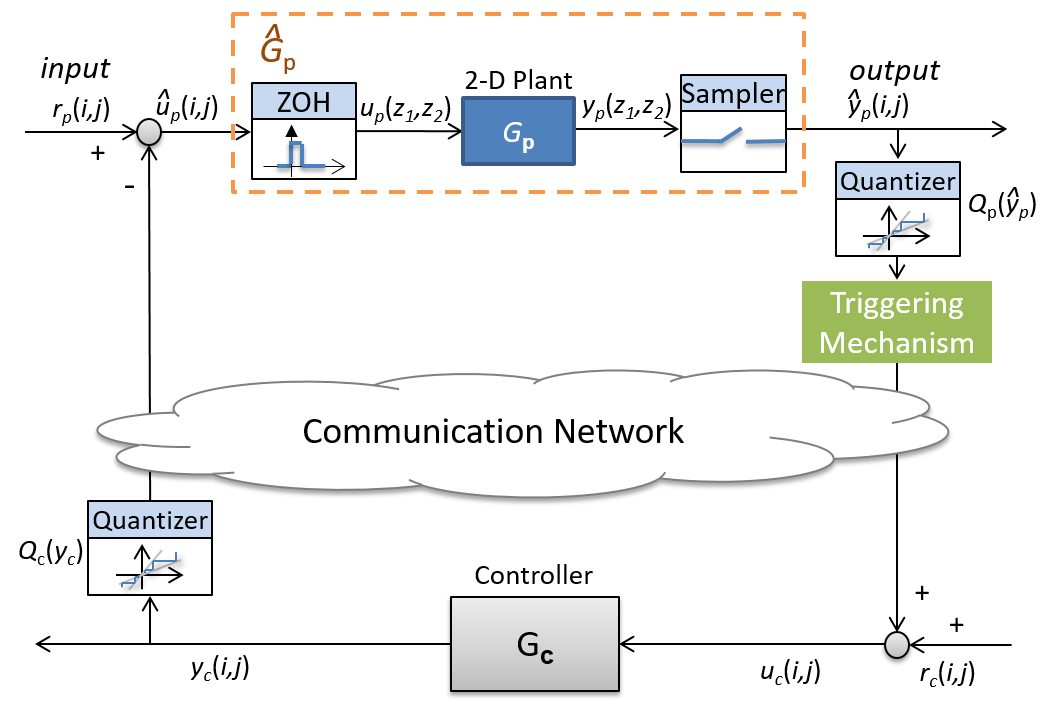}
 \caption{A 2-D system Interconnected with a Controller over a Digital Network}
 \label{fig:2DNCS}
\end{figure}
%

\section{Main Results}
\label{sec:main results}
In this section, we present the main results of the paper. We consider the three operations -- sampling, quantization, and event triggered transmission -- in a digital network sequentially. The proofs of all the technical results can be found in the appendix.

\subsection{Dissipassivity of Sampled 2-D Systems}
\label{sec:2-D Approximation}
We begin by considering the impact of sampling. Thus, we focus on the system $\hat{G}_{p}$ depicted in the dashed box in Figure~\ref{fig:2DNCS} whose output is obtained by sampling the output of the continuous system $G_{p}$ and whose input is applied to $G_{p}$ through a zero order hold. 

Let us start with linear systems.
We consider the continuous 2-D Roesser linear model from ({\ref{eq:2-DCTsystem}}) and we assume that $A_{11}\in\mathbb{R}^{n_{h}\times n_{h}}$ and $A_{22}\in\mathbb{R}^{n_{v}\times n_{v}}$ are nonsingular for simplicity. By considering a storage function $V\left(x_h(i,j),x_v(i,j)\right)=x^T_h(i,j)P_hx_h(i,j)+x_v^T(i,j)P_vx_v(i,j)$, the following result provides a sufficient condition for establishing the QSR-dissipativity of a sampled system.

\begin{theorem}
\label{thm:linear passivity index}
Consider a sampled 2-D model with sampling period $h_{1},h_{2}$. The sampled system is $(\hat{Q}_p,\hat{S}_p,\hat{R}_p)$-dissipative if there exist matrices $P_{h}>0$ and $P_{v}>0$ such that
\begin{align}
\label{eq:$QSR$ LMI}
\begin{split}
\begin{pmatrix}A_{d}^{T}PA_{d}-P-C_{d}^{T}\hat{Q}_pC_{d} & *\\
A_{d}^{T}PB_{d}-C_{d}^{T}\hat{Q}_pD-C^{T}\hat{S}_p & \Phi
\end{pmatrix}\le0,
\end{split}
\end{align}
where 
\begin{equation}
\left\{ \begin{array}{l}
    \Phi= B_{d}^{T}PB_{d}-\hat{R}_p-D_{d}^{T}\hat{Q}_pD_{d}-\hat{S}_p^{T}D_{d}-D_{d}^T\hat{S}_p\\
    A_{d}=\begin{pmatrix}
        e^{A_{11}h_{1}} & (e^{A_{11}h_{1}}-I)A_{11}^{-1}A_{12}\\
        (e^{A_{22}h_{2}}-I)A_{22}^{-1}A_{21} & e^{A_{22}h_{2}}
    \end{pmatrix}\\
    B_{d}=\left(\begin{array}{c}
        (e^{A_{11}h_{1}}-I)A_{11}^{-1}B_{1}\\
        (e^{A_{22}h_{2}}-I)A_{22}^{-1}B_{2}
    \end{array}\right)\\
    C_{d}=\left(\begin{array}{cc}
        C_{1} & C_{2}
    \end{array}\right),
    D_{d}=D\\
    P=\begin{pmatrix}
        P_{h} & 0 \\ 0 & P_{v}
    \end{pmatrix}.
\end{array}
\right.
\label{eq:ABCD DT}
\end{equation}
\end{theorem}

\begin{remark}
    In the above theorem, it is assumed for simplicity that $A_{11}$ and $A_{22}$ are nonsingular. Whenever $A_{11}$ or $A_{22}$ is a singular matrix, the matrix-valued function $(e^{A_{11}h_1}-I)A_{11}^{-1}$, and $(e^{A_{22}h_2}-I)A_{22}^{-1}$ respectively, in the presentation of Theorem 2 are represented as $\Sigma_{k=1}^{\infty} h_1(A_{11}h_1)(k-1)/k$, and $\Sigma_{k=1}^{\infty} h_2(A_{22}h_2)(k-1)/k$, respectively, {\cite{Chen99}}. Note that the summation of the series can be obtained by numerical methods, see for instance {\cite{Nakamura90}}.
\end{remark}
%

Next, let us consider the nonlinear case. We state the following assumption that is made following \cite{Oishi10} and \cite{Xia17} which study the impact of sampling on disspitavity of 1-D systems.
\begin{assumption}
\label{asmp: good approximation}
    There exist scalars $\alpha_1,\alpha_2 \geq 0$ such that for all $Z_1 \geq 0$, $Z_2 \geq 0$ and for all $u_p \in \mathcal{U}$,
\begin{align}
    \label{eq:2-D dot_dot_y<a||u||^2}
    \begin{split}
        &\int^{Z_1}_0 \left|\frac{\partial y_p(z_1,z_2)}{\partial z_1}\right|^2 dz_1
        \leq \alpha_1^2 \int^{Z_1}_0 \left|u_p(z_1,z_2)\right|^2 dz_1, \\
        &\int^{Z_2}_0  \left|\frac{\partial y_p(z_1,z_2)}{\partial z_2}\right|^2 dz_2 
        \leq  \alpha_2^2 \int^{Z_2}_0 \left|u_p(z_1,z_2)\right|^2 dz_2.
    \end{split}
\end{align}
\end{assumption}

This assumption puts a constraint on how fast the output response can change, with respect to each direction, for all admissible inputs. Note that Section 5.3 in {\cite{Khalil00}} gives us some insights of how to find such $\alpha_1$ and $\alpha_2$ for a given 2-D system. They can be viewed as the $\mathcal{L}_2$ gain of the mappings $u(z_1,z_2)\rightarrow \frac{\partial y_p(z_1,z_2)}{\partial z_1}$ and $u(z_1,z_2)\rightarrow \frac{\partial y_p(z_1,z_2)}{\partial z_2}$ respectively.

Denote by $h_1$ and $h_2$ the sampling period for the horizontal and vertical coordinates in 2-D domains, respectively. 
Denote the input (respectively the output) of $\hat{G}_p$ as $\hat{u}_p(ih_1,jh_2)$ (respectively $\hat{y}_p(ih_1,jh_2)$),  which is abbreviated for notational convenience as $\hat{u}_{p}(i,j)$ (respectively $\hat{y}_{p}(i,j)$) for the rest of this paper. 

Thus,
\begin{equation}
\label{eq:ZOH}
\left.\begin{array}{l}
u_p(z_1,z_2)=\hat{u}_p(i,j)\\
\hat{y}_p(i,j)=y_p(ih_1,jh_2)\\
\end{array}\right\}   \begin{array}{c}
\forall z_1\in[ih_{1},(i+1)h_{1})\\
\forall z_2\in[jh_{2},(j+1)h_{2})
\end{array},
\end{equation}
for $\forall i,j\in\mathbb{N}$. 

Denote by $\Delta y_p(z_1,z_2)$ the difference between outputs of $G_p$ and $\hat{G}_p$, i.e.,
\begin{align}
    \label{eq:2-D delta y}
    \Delta y_p(z_1,z_2)=y_p(z_1,z_2)-\hat{y}_p(i,j),
\end{align}
where $z_1\in\lbrack ih_1, (i+1)h_1)$ and $z_2\in\lbrack jh_2,(j+1)h_2)$ for all $i,j\in\mathbb{N}$.

The following result reveals the relation between sampling error $\Delta y_p$ and sampling periods, $h_1$, $h_2$.
\begin{lemma}
\label{lem:2-D delta_y^2<ru^2}
Consider a continuous 2-D system $G_p$ and its sampled model $\hat{G}_p$  as shown in Figure \ref{fig:2DNCS}. Under Assumption \ref{asmp: good approximation}, for all $\forall N_1,N_2 \in \mathbb{N}$ with $Z_1=N_1h_1$, $Z_2=N_2h_2$, we have
\begin{align}
    \label{eq:2-D (delta_y)^2<a(u^2)}
    \begin{split}
&\int^{Z_1}_0 \int^{Z_2}_0 \left|\Delta y_p(z_1,z_2)\right|^2 \text{ }dz_2dz_1 \\ 
& \leq 2(\alpha_1^2h_1^2+\alpha_2^2h_2^2) \int^{Z_1}_0 \int^{Z_2}_0 \left|u_p(z_1,z_2)\right|^2 dz_2dz_1.
\end{split}
\end{align}
\end{lemma}

Note that it is difficult to obtain the relation between sampling error $\Delta y_p(z_1,z_2)$ and sampling periods $h_1$ \& $h_2$, in 2-D domains. Although it may be easy to derive a simplified result (by eliminating either dimension) for 1-D systems, extending to 2-D systems is non-trivial, see the proof of Appendix-B for details.

The following result relates the $QSR$-dissipativity of the continuous nonlinear 2-D system $G_{P}$ and its corresponding sampled 2-D model $\hat{G}_{p}$.

\begin{theorem}
\label{thm:2-D passivity index}
Consider a $(Q_p,S_p,R_p)$-dissippative continuous nonlinear 2-D system $G_p$ and obtain a discrete system $\hat{G}_p$ using sampling and zero-order hold as shown in Figure \ref{fig:2DNCS}. Under Assumption \ref{asmp: good approximation}, $\hat{G}_p$ is $(\hat{Q}_p,\hat{S}_p,\hat{R}_p)$-disspative if there exist constants $\xi_1,\xi_2,\xi_3\in \mathbb{R}^{+}$ such that 
\begin{equation}
\begin{array}{rcl}
\label{eq:2-D passivity index}
\underline{\lambda}(\hat{Q}_{p}-Q_{p}) & \ge & \xi_{1}\bar{\sigma}_{max}^{2}(\hat{Q}_{p})+\xi_{2}\bar{\sigma}_{max}^{2}(\hat{S}_{p}-S_{p}),\\
\underline{\lambda}(\hat{R}_{p}-R_{p}) & \ge & 2(\alpha_{1}h_{1}+\alpha_{2}h_{2})(|\underline{\lambda}(\hat{Q}_{p})|+\frac{1}{\xi_{1}}+\frac{1}{\xi_{3}})\\
 &  & +\xi_{3}\bar{\sigma}_{max}^{2}(\hat{S}_{p})+\frac{1}{\xi_{2}}.
\end{array}
\end{equation}
\end{theorem}

Theorem 1 proposes a method to estimate the $QSR$-dissipativity of a sampled system using the dissipativity matrices for the original system and the sampling periods. 

\subsection{Quantization of Sampled 2-D Systems}
\label{sec:quantization}
In the rest of this work, we assume the dimensions of the input and output are the same, i.e., $m=p$, and we restrict the plant and controller to be IF-OFP systems. The main reason that we consider IF-OFP systems is that it enables us to get a more clear and verifiable condition and IF-OFP levels provide a quantifiable measure of the dissipativity degradation over network induced effects.

\begin{figure}
\centering
 \captionsetup{font={scriptsize}}
 \includegraphics[scale = 0.58]{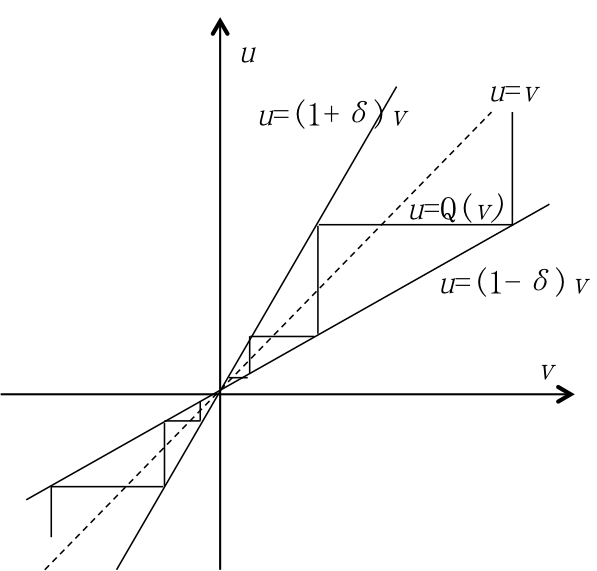}
 \caption{Logarithmic Quantizer}
 \label{fig:quantizer}
\end{figure}

We now consider the impact of quantization on dissipativity. Specifically, we consider the framework shown in Figure~\ref{fig:2DNCS}, but without the event-triggering mechanism present. The outputs of the two systems, i.e., $\hat{y}_p(i,j)$ and $y_c(i,j)$, are quantized using quantizers $\mathcal{Q}_p$ and $\mathcal{Q}_c$, respectively, before transmission. 
We focus on static logarithmic quantizers that have been utilized widely in networked control systems (see, e.g., \cite{Elia01} \cite{fu2005sector} \cite{coutinho2010input}). As illustrated in Figure \ref{fig:quantizer}, this quantizer is defined as 
\begin{align}
\label{eq:2-D quantier}
\mathcal{Q}(v)=\left\{ \begin{array}{ll}
\theta^{i} & \text{if}\;\frac{1}{1+\delta}\theta^{i}<v\leq\frac{1}{1-\delta}\theta^{i}\\
& v>0,\;i=\pm1,\pm2,\ldots\\
0 & \text{if}\;v=0\\
-\mathcal{Q}(-v) & \text{if}\;v<0
\end{array}\right.,
\end{align}
where $v\in\mathbb{R}$ is the input and $\mathcal{Q}(v)$  is the output of the quantizer, $0<\theta<1$ represents the quantization density, $\delta=\frac{1-\theta}{1+\theta}$ and $\delta\in(0,1)$. Thus, the quantization error, defined as $\Delta v\triangleq\mathcal{Q}(v)-v$, admits the following bound condition 

\begin{align}\label{eq:bound_quantizer}
|\Delta v|\leq\delta|v|,
\end{align}
which can be equivalently rewritten as
\begin{align}
    (1-\delta)v^{2}\leq v\mathcal{Q}(v)\leq(1+\delta)v^{2}.
\end{align}
 We assume (following \cite{Zhu12}) that $\mathcal{Q}(\cdot)$ acts as a component-wise operator on the input signal if the input to the quantizer is a vector.

The following result characterizes the effect of quantization on  $\mathcal{L}_2$-stability of the feedback interconnected 2-D system based on the passivity levels. 
\begin{theorem}
\label{thm:2-D passivity index with quantization}
Consider a IF-OFP($\rho_p,\nu_p$) 2-D system $\hat{G}_p$ and a IF-OFP($\rho_c,\nu_c$) 2-D system $G_c$  interconnected through a feedback loop with logarithmic quantizers at the outputs as shown in Figure \ref{fig:2DNCS} (but with the event triggered communication absent). The closed-loop map is $\mathcal{L}_2$-stable if there exist $\beta_1 ,\beta_2\in\mathbb{R}^+$ such that
\begin{align}
    \label{eq:2-D passivity index with quantization 1}
        &\rho_{p}+\nu_{c}>(\delta_{p}^{2}+2\delta_{p})|\nu_{c}|+(1+\frac{\beta_{2}}{2})\delta_{p}^{2}+\frac{1}{2\beta_{1}},
 \\
    \label{eq:2-D passivity index with quantization 2}
        &\rho_{c}+\nu_{p}>(\delta_{c}^{2}+2\delta_{c})|\nu_{p}|+(1+\frac{\beta_{1}}{2})\delta_{c}^{2}+\frac{1}{2\beta_{2}}.
\end{align}
\end{theorem}
%


%

\begin{remark}
Theorem {\ref{thm:2-D passivity index with quantization}} reduces to Lemma {\ref{lem:2-D closed loop stable using index}} when no quantizer is adopted in network while the right hand side of above inequalities can be interpreted as the deficiency of passivity caused by quantization. 
\end{remark}

Note that Theorem \ref{thm:2-D passivity index with quantization} does not assume that both the components are passive. In case one of the components (say system $\hat{G}_p$) has a shortage of passivity, i.e. with negative passivity levels, the other component (say $G_c$) can be used to compensate this shortage as well as the loss of passivity caused by quantization. The following remark gives us a design-oriented perspective for such controllers.

\begin{remark}
\label{thm:2-D close loop stability}
Consider a IF-OFP($\rho_p,\nu_p$) 2-D system $\hat{G}_p$ and a IF-OFP($\rho_c,\nu_c$) 2-D system $G_c$  interconnected through a feedback loop with logarithmic quantizers at their outputs as shown in Figure \ref{fig:2DNCS} (but with the event-triggered communication absent).  
If the controller satisfies
\begin{align}
\label{eq:2-D $QSR$ controller1}
    &\nu_c-(\delta_p^2+2\delta_p)\left|\nu_c\right| > -\rho_p+(1+\frac{\beta_2}{2})\delta_p^2+\frac{1}{2\beta_1},\\
\label{eq:2-D $QSR$ controller2}
    &\rho_c >-\nu_p+(\delta_c^2+2\delta_c)\left|\nu_p\right|+(1+\frac{\beta_1}{2})\delta_c^2+\frac{1}{2\beta_2},
\end{align}
where $\beta_1,\beta_2 \in \mathbb{R}^{+}$, then the feedback interconnected system is $\mathcal{L}_2$-stable.
\end{remark}

In this setting, the representation-free manner of controller design enhances the closed-loop performance from a practical point of view. Since the passivity levels of a system can be changed using passivation method (see, for instance, \cite{Xia17} \cite{Bao07}), the constants $\beta_1$ and $\beta_2$ in (\ref{eq:2-D $QSR$ controller1})-(\ref{eq:2-D $QSR$ controller2}) leave a lot of room for the designer to design the controller in order to ensure stability. 

It can be observed from (\ref{eq:2-D quantier})-(\ref{eq:bound_quantizer}) that the logarithmic quantizer has an infinite number of discrete quantization levels due to its time-invariance nature, which makes it impractical to be implemented. In practice, it is usually needed to set a small enough value $\epsilon>0$, which is called dead zone, and let $Q(v)=0$, when $|v|<\epsilon$. In this way, a finite quantization levels are available given bounded input, and the practical stability can be achieved. 

%
%
%
%
%

%

\subsection{Event-triggered Design of Quantized 2-D Systems}
\label{sec:2-D event}
We now consider the complete digital control framework shown in Figure~\ref{fig:2DNCS}. Now it requires a system to only broadcast its quantized output information when the local output error exceeds a given threshold. It is worth mentioning that in 2-D domains, there exist two independent coordinates so the local output error is spread over a quadrant. This fact definitely leads to technical challenge for the event-triggering design. Note that the following assumption is common in practice since a wide range of physical applications have bounded domains. For instance, the spacial domain of thermal processes modeled by 2-D systems is bounded \cite{Marszalek84}, and both the horizontal and vertical domains of 2-D image processing are bounded \cite{sonka2014image}.

\begin{assumption}
We restrict our attention to 2-D systems with one of the coordinates varies on a bounded spatial domain.
\end{assumption}

As it will be explained in Remark \ref{remark:event infinite}, the proposed methodology can be used also for the general 2-D systems with infinite horizontal and vertical coordinates. For clarity of exposition, we assume that the horizontal coordinate represents the space which is bounded, while the vertical coordinate represents time. Before presenting our main result of this subsection, let us introduce the following definition.

\begin{definition}
\label{def: bounded l2 -2d}
A nonlinear discrete 2-D system (\ref{eq:2-D discrete nonlinear system}) with bounded spatial domain, i.e., $i\in[0,N_{1}-1]$, is said to have finite $\mathcal{L}_{2}$-gain if there exist constants $\gamma> 0,\zeta$ such that for all $N_{2}\in\mathbb{N}$, all $x_{h}(0,j)\in\mathcal{X}_{h}$, $x_{v}(i,0)\in\mathcal{X}_{v}$, and all input function $u\in\mathcal{U}$, 
\begin{align}
\label{eq:bounded l2 - 2d}
\sum_{i=0}^{N_{1}-1}\sum_{j=0}^{N_{2}-1}|y(i,j)|^{2}\le\gamma\sum_{i=0}^{N_{1}-1}\sum_{j=0}^{N_{2}-1}|u(i,j)|^{2}+\zeta.
\end{align}
\end{definition}

Event-triggered transmission has become very popular in networked control literature, see for instance \cite{peng2013event} \cite{peng2015event} \cite{eqtami2010event} \cite{wang2011event}. In this work, we will design a triggering mechanism to determine the transmission instants from the system $\hat{G}_{p}$ to $G_{c}$. Denote by $j_{k},k=1,2,\ldots$ the $k$-th triggering instants. The following theorem provides such an event-triggered mechanism that preserves $\mathcal{L}_{2}$-stability.

\begin{theorem}
\label{thm: 2-D triggering}
Consider the framework in Figure \ref{fig:2DNCS}, wherein the 2-D systems  $\hat{G}_{p}$ and $G_{c}$ have bounded spatial domains and they are IF-OFP$(\rho_{p},\nu_{p})$ and IF-OFP$(\rho_{c},\nu_{c})$, respectively. Suppose that there exist $\beta_{1},\beta_{2}\in\mathbb{R}^{+}$ such that $q_{1},q_{2}<0$ where 
\[
\left\{ \begin{array}{c}
q_{1}=-(\rho_{p}+\nu_{c})+(\delta_{p}^{2}+2\delta_{p})|\nu_{c}|+(\frac{\beta_{2}}{2}+1)\delta_{p}^{2}+\frac{1}{2\beta_{1}}\\
q_{2}=-(\rho_{c}+\nu_{p})+(\delta_{c}^{2}+2\delta_{c})|\nu_{c}|+(\frac{\beta_{1}}{2}+1)\delta_{p}^{2}+\frac{1}{2\beta_{2}}.
\end{array}\right.
\]
Assume for each $k=1,2, \cdots,$ the next triggering instant $j_{k+1}$
is chosen according to
\begin{equation}
\begin{array}{rl}
j_{k+1}= & \underset{{\scriptstyle j\in\mathbb{N},j>j_{k}}}{\text{min}}  j\\
\text{s.t.} & {\displaystyle \sum_{i=0}^{N_{1}-1}}\left|\epsilon e(i,j)+\frac{\nu_{c}}{\epsilon}\mathcal{Q}_{p}(\hat{y}_{p}(i,j))\right|^{2}>\\
 & \left(\frac{\nu_{c}^{2}}{\epsilon^{2}}-\frac{\theta_{1}q_{1}}{(1+\delta_{p})^{2}}\right){\displaystyle \sum_{i=0}^{N_{1}-1}}\left|\mathcal{Q}_{p}(\hat{y}_{p}(i,j))\right|^{2},
\end{array}\label{eq:triggering condition}
\end{equation}
where $e(i,j)\triangleq\mathcal{Q}_{p}(\hat{y}_{p}(i,j))-\mathcal{Q}_{p}(\hat{y}_{p}(i,j_{k}))$
and $\epsilon^{2}=|\nu_{c}|-\nu_{c}-\frac{1}{4\theta_{2}q_{2}}$ with
$\theta_{1},\theta_{2}\in(0,1)$. Then the interconnected system has
a finite $\mathcal{L}_{2}$-gain. 
\end{theorem}

Under the event-triggering mechanism, the sensor connected to the plant output determines when to transmit the sampled output to the network in order to guarantee the closed-loop stability, which can be summarized as follows. First, we set the initial triggering instant as $j_{1}=0$. Specifically, at the initial time step, the quantized output of the sampled system $\hat{G}_{p}$ at all positions on horizontal coordinate, i.e., $\mathcal{Q}_{p}(\hat{y}_{p}(i,0)),i\in\{0,\ldots,N_{1}-1\}$ will be sent to the controller $G_{c}$. Next, at each $k=1,2\ldots$, we examine the condition \eqref{eq:triggering condition} at each time step $j\ge j_{k}+1$. If the triggering condition \eqref{eq:triggering condition} is not satisfied at certain time step, no update will be implemented by the transmitter at this moment, but $j$ increases by 1. In this case, the input of the controller $G_{c}$ is given by $u_{c}(i,j)=r_{c}(i,j)+\mathcal{Q}_{p}(\hat{y}_{p}(i,j_{k}))$. Whenever the triggering condition \eqref{eq:triggering condition} holds, the quantized output of the sampled system $\hat{G}_{p}$ at all positions on the horizontal coordinate, i.e., $\mathcal{Q}_{p}(\hat{y}_{p}(i,j)),i\in\{0,\ldots,N_{1}-1\}$ will be updated and transmitted to the controller $G_{c}$. At the same time, we update $j_{k+1}=j$. We repeat this process until the end. It is worth mentioning that the proposed triggering mechanism can also be applied to the general 2-D domains. For instance, when $i$ and $j$ are both spatial coordinates representing horizontal and vertical directions, the proposed triggering mechanism can be exploited to determine the necessary information exchanging moments alongside the horizontal coordinate or the vertical coordinate. 

\begin{remark}
\label{remark:event infinite}
The event-triggering condition in Theorem \ref{thm: 2-D triggering} is obtained by comparing the output errors accumulated in finite spatial domain along the temporal coordinate. This methodology can be extended to general 2-D systems with infinite spatial domains. Specifically, one can grid the infinite spatial domain into consecutive intervals, i.e., $i\in[0,N_1-1],[N_1,2N_1-1],\cdots,$ and then perform the event-triggering mechanism for each interval. In this way, the condition in (\ref{eq:bounded l2 - 2d}) is satisfied for the sum over $i\in[0,N_1-1],[N_1,2N_1-1],\cdots$. This implies that the discrete 2-D system has a finite $\mathcal{L}_2$ gain according to Definition \ref{def:passivity}.  
\end{remark}


\section{Simulation and Example}
\label{sec:2-D example}

Consider the hyperbolic partial differential equation (PDE) of second order \cite{Marszalek84}, which is described as 
\begin{equation}
\frac{\partial^{2}s(x,t)}{\partial x\partial t}=a_{1}\frac{\partial s(x,t)}{\partial t}+a_{2}\frac{\partial s(x,t)}{\partial x}+a_{0}s(x,t)+bf(x,t),\label{eq:pde}
\end{equation}
with the boundary conditions 
\[
s(0,t)=q(t),\; s(x,0)=p(x),
\]
where $a_{1},a_{2},a_{0},b$ are real constants and $f(x,t)$ is the system input. Such a class of equations can be used to describe linear processes in thermal engineering such as gas absorption and water stream heating.

Taking the feedback control of a heat exchanger as a practical example, here, $s(x,t)$ represents the temperature at $x$(space) and $t$(time). $f(x,t)$ is a given force function. By defining $r(x,t)\triangleq\frac{\partial s(x,t)}{\partial t}-a_{2}s(x,t)$ as an internal variable, we can transform the hyperbolic PDE into a 2-D continuous Roesser model given by 
\begin{equation}
\left[\begin{array}{c}
\frac{\partial r(x,t)}{\partial x}\\
\frac{\partial s(x,t)}{\partial t}
\end{array}\right]=\left[\begin{array}{cc}
a_{1} & a_{1}a_{2}+a_{0}\\
1 & a_{2}
\end{array}\right]\left[\begin{array}{c}
r(x,t)\\
s(x,t)
\end{array}\right]+\left[\begin{array}{c}
b\\
0
\end{array}\right]f(x,t)\label{eq:example}
\end{equation}
with the boundary conditions 
\begin{equation}
\left\{ \begin{array}{rc}
r(0,t)= & \frac{dq(t)}{dt}-a_{2}q(t),\\
s(x,0)= & p(x).
\end{array}\right.\label{eq:boundary_example}
\end{equation}

We consider the above PDE equation with $a_{0}=1$, $a_{1}=1$, $a_{2}=-1$, $b=1$,
$q(t)=1$, $p(x)=e^{-x}$, and consider the output as $y(x,t)=r(x,t)+s(x,t)$, which is the changing rate of the temperature. By taking $r(x,t)$ as horizontal state and $s(x,t)$ as vertical state, then the continuous Roesser model \eqref{eq:2-DCTsystem} under consideration is given
by
\begin{equation}
\left\{ \begin{array}{l}
\left[\begin{matrix}A_{11} & A_{12}\\
A_{21} & A_{22}
\end{matrix}\right]=\left[\begin{matrix}1 & 0\\
1 & -1
\end{matrix}\right],\text{ }\left[\begin{matrix}B_{1}\\
B_{2}
\end{matrix}\right]=\left[\begin{matrix}1\\
0
\end{matrix}\right],\\
\begin{bmatrix}C_{1} & C_{2}\end{bmatrix}=\begin{bmatrix}1 & 1\end{bmatrix},\text{ }D=\emptyset,
\end{array}\right.\label{eq:Roesser_example}
\end{equation}
with the boundary condition $r(0,t)=1$, $s(x,0)=e^{-x}$. 

To begin with, let us observe that the 2-D continuous system \eqref{eq:Roesser_example} is unstable since we have (See Theorem 4 in \cite{Agathoklis91} for details)
\begin{align*}
    \text{Re}[\lambda(A_{11})]=1>0.
\end{align*}

The problem is to establish whether the closed-loop system with a given output feedback controller $G_{c}$ over a digital network as depicted in Figure \ref{fig:2DNCS} is stable.

\begin{figure}
\centering
\includegraphics[scale=0.4]{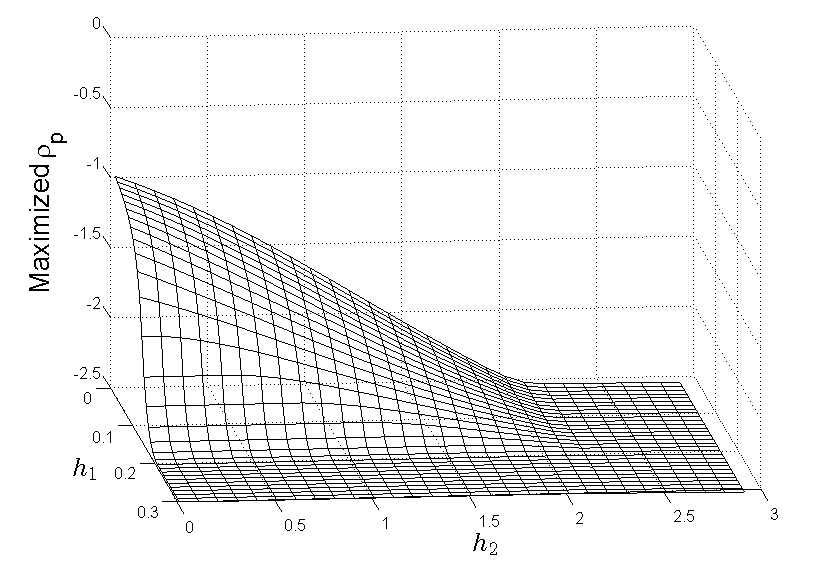}
\captionsetup{justification=centering}
\caption{Maximized $\rho_p$ obtained via Theorem 2}
\label{fig:ex_1}
\end{figure}
\begin{figure}
\centering
\includegraphics[scale=0.37]{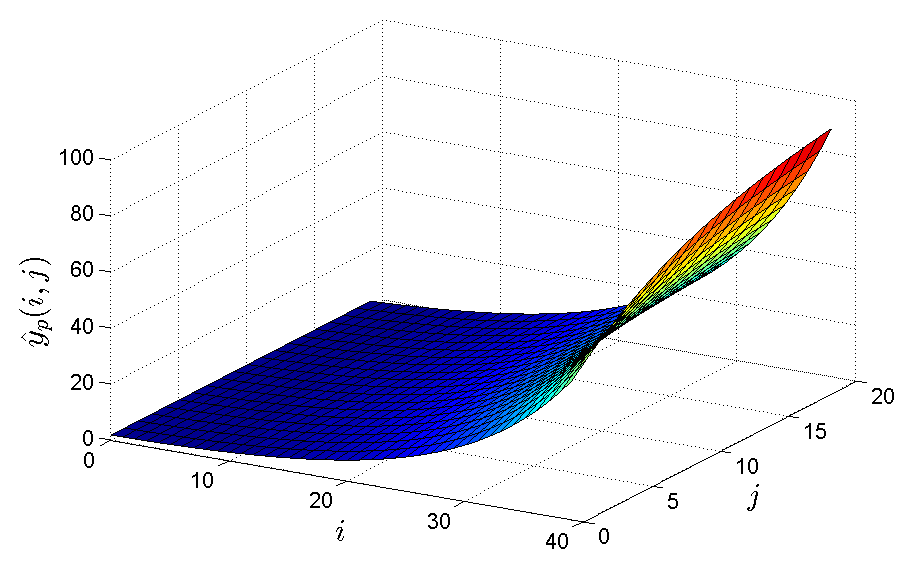}
\caption{The trajectory of output $\hat{y}_p(i,j)$}
\label{fig:ex_2}
\end{figure}

Firstly, let us estimate the $(\hat{Q}_{p},\hat{S}_{p},\hat{R}_{p})$-dissipativity of the sampled system $\hat{G}_{p}$ corresponding to the continuous 2-D system \eqref{eq:Roesser_example}. Since the system $G_p$ described in \eqref{eq:Roesser_example} has single input and single output, the $(\hat{Q}_{p},\hat{S}_{p},\hat{R}_{p})$-dissipativity of its sampled system $\hat{G}_{p}$ can be equivalently expressed by its IF-OFP levels $(\rho_{p},\nu_{p})$ by letting $\hat{Q}_{p}=-\rho_{p},\hat{S}_{p}=0.5,\hat{R}_{p}=-\nu_{p}$. Given the sampling periods $h_{1},h_{2}$ and a fixed $\nu_{p}$ (respectively $\rho_{p}$), the lower bound of maximum $\rho_{p}$ (respectively $\nu_{p}$) can be derived via maximizing $\rho_{p}$ (respectively $\nu_{p}$) subject to the LMI condition proposed in (\ref{eq:$QSR$ LMI}). Here, we fix $\nu_{p}=-0.1$, and compute the maximized $\rho_{p}$ for different values of $(h_{1},h_{2})\in[0,0.3]\times[0,3]$ based on (\ref{eq:$QSR$ LMI})-(\ref{eq:ABCD DT}), which is plotted in Figure \ref{fig:ex_1}. As shown in Figure \ref{fig:ex_1}, with fixed $h_{1}$(respectively $h_{2}$), the maximized $\rho_{p}$ decreases monotonically as $h_{2}$ (respectively $h_{1}$) increasing, and it remains constant when it hits the value $\rho_{p}=-2.5$ resulting from the fact that the domain of $(\rho,\nu)$ of a IF-OFP system is constrained as stated in Lemma \ref{lem:2-D passivity index <1/4}.

With $h_{1}=h_{2}=0.1$
and $\nu_{p}=-0.1$, we obtain $\rho_{p}=-1.317$. This set of
values will be used in the following steps. 

With $h_{1}=h_{2}=0.1$, we compute the output of the sampled system $\hat{y}_p(i,j)$ under the boundary conditions \eqref{eq:boundary_example}. It is shown by Figure \ref{fig:ex_2} that the sampled system is unstable since $\hat{y}_p(i,j)$ diverges as $i$ or $j$ increasing. 
\begin{figure*}
\centering
\includegraphics[scale=0.55]{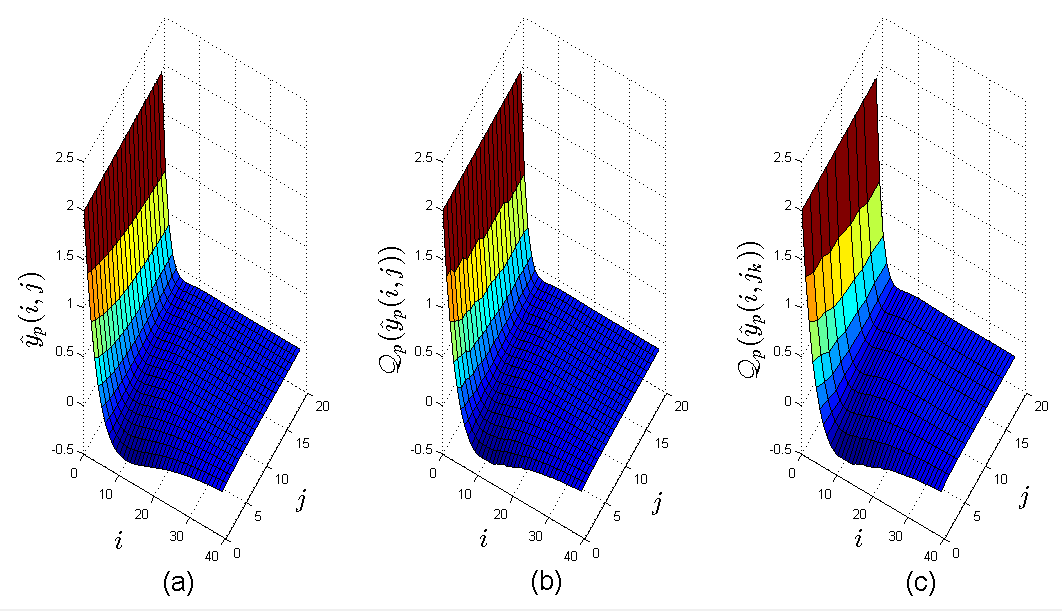}
\caption{Trajectories of transmitted output of $\hat{G}_p$: (a)without digital network; (b) with quantizers; (c) with quantizers and event triggered communication}
\label{fig:ex_3}
\end{figure*}

Next, with $h_{1}=h_{2}=0.1$, we consider the closed-loop system in Figure \ref{fig:2DNCS} with event-triggered communication absent. It is assumed that the output of the sampled system $\hat{y}_p(i,j)$ and the output of the controller $y_{c}(i,j)$ are quantized before being transmitted by the logarithmic quantizer $\mathcal{Q}_{p}(\cdot)$ and $\mathcal{Q}_{c}(\cdot)$, respectively, saying $\delta_{c}=\delta_{p}=0.04$ in this example. The first task is searching for a controller $G_{c}$ with IF-OFP levels $(\rho_{c},\nu_{c})$ satisfying (\ref{eq:2-D $QSR$ controller1})-(\ref{eq:2-D $QSR$ controller2}) with some $\beta_{1},\beta_{2}\in\mathbb{R}^{+}$. In this example, we consider a static linear output feedback controller $y_{c}(i,j)=Ku_{c}(i,j)$. Specifically, given $\rho_{p}=-1.317$, $\nu_{p}=-0.1$ and $\delta_{c}=\delta_{p}=0.04$, it can be easily obtained that by letting $K=3$, the feedback controller $G_{c}$ is IF-OFP with $\rho_{c}=0.166,\nu_{c}=1.5$ while the conditions in (\ref{eq:2-D $QSR$ controller1})-(\ref{eq:2-D $QSR$ controller2}) being satisfied with $\beta_{1}=36$, $\beta_{2}=56$.
Figure \ref{fig:ex_3} (a) shows the closed-loop response $\hat{y}_{p}(i,j)$ of the sampled system without quantization, i.e., $\delta_{c}=\delta_{p}=0$, while Figure \ref{fig:ex_3} (b) shows the trajectory of the quantized output of the sampled system with $\mathcal{Q}_{p}(\hat{y}_{p}(i,j))$. It is shown in Figure \ref{fig:ex_3} (b) that with the feedback controller, the trajectory of the system output under quantization converges to zero as $i$ and $j$ increase. 

\begin{figure}
\centering
\includegraphics[scale=0.32]{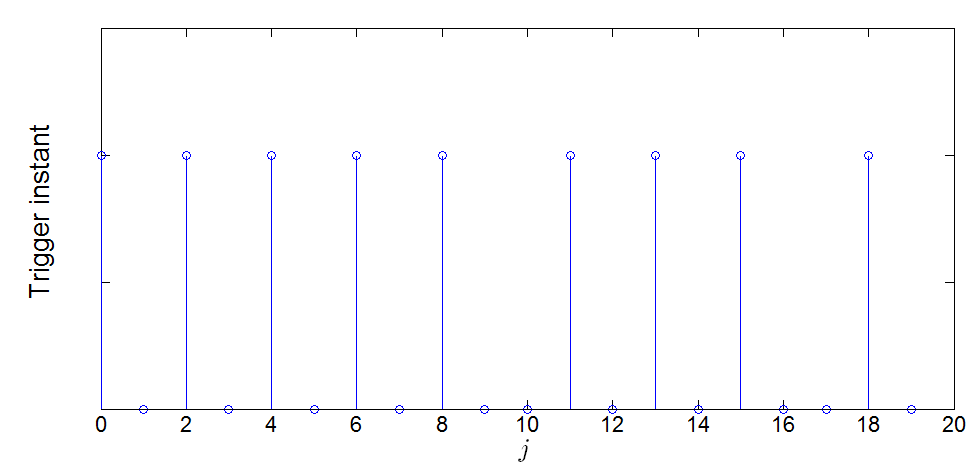}
\captionsetup{justification=centering}
\caption{Triggering instants along $j$ coordinate }
\label{fig:ex_4}
\end{figure}

In the end, we consider the complete framework in Figure \ref{fig:2DNCS} by applying the triggering mechanism proposed in Theorem \ref{thm: 2-D triggering}. Specifically, we assume that the spacial domain is bounded by $N_{1}=40$, i.e., $i\in[0,39]$. Let us set $\theta_{1}=\theta_{2}=0.5$. Based on Theorem \ref{thm: 2-D triggering}, the quantizted output of the sampled system $\hat{G}_{p}$, $\mathcal{Q}_{p}(\hat{y}_{p}(i,j)),i\in\{0,1,\ldots,39\}$ is transmitted only when the triggering condition in (\ref{eq:triggering condition}) is satisfied. Figure \ref{fig:ex_4} shows the triggering instants along the $j$ coordinate while Figure \ref{fig:ex_3} (c) plots the trajectory of transmitted signal $\mathcal{Q}_{p}(\hat{y}_{p}(i,j_{k}))$. It is shown by Figure \ref{fig:ex_3} (c) that the unstable 2-D continuous system \eqref{eq:Roesser_example} is stable under feedback interconnection over the digital network as depicted in Figure \ref{fig:2DNCS}.

\section{Conclusion}
\label{sec:2-D conclusion}
In this paper, we have studied thedissipativity analysis for a continuous 2-D system across a digital communication network. First, dissipativity of sampled nonlinear 2-D systems is characterized. Then, the effects of signal quantization in communication links on dissipativity of such systems is considered. In the end, an event-triggered scheme is proposed for 2-D networked control systems. Based on these conditions, the $\mathcal{L}_2$-stability and robustness of the closed-loop system is guaranteed under these network induced effects.

\section*{Appendix}
\subsection{Domain of Passivity Indices}
\begin{lemma}
\label{lem:2-D passivity index <1/4}
\cite{Matiakis06}
    The range of $\rho,\nu$ in a IF-OFP system is $\Omega=\Omega_1\cup\Omega_2$, with $\Omega_1=\left\{\rho,\nu\in\mathbb{R}|\rho \nu<\frac{1}{4}\right\}$ and $\Omega_2=\left\{\rho,\nu\in\mathbb{R}|\rho\nu=\frac{1}{4},\rho>0\right\}$.
\end{lemma}

\subsection{Proof of Lemma \ref{lem:2-D delta_y^2<ru^2}}

\begin{IEEEproof}
First, it can be obtained that for all $z_1\in[ih_1,(i+1)h_1)$, $z_2\in[jh_2,(j+1)h_2)$ and all $i,j\in\mathbb{N}$,
\begin{equation}
\begin{array}{rl}\label{eq:2-D caucgy scgward}
    &\left| \int^{z_1}_{ih_1} \frac{\partial y_{p}(s_1,z_2)}{\partial s_1}ds_1 +\int^{z_2}_{jh_2}\frac{\partial y_{p}(ih_1,s_2)}{\partial s_2} ds_2 \right|^2 \\
    \leq &2
    \left| \int^{z_1}_{ih_1} \frac{\partial y_{p}(s_1,z_2)}{\partial s_1} ds_1 \right|^2 + 2 \left| \int^{z_2}_{jh_2} \frac{\partial y_{p}(ih_1,s_2)}{\partial s_2} ds_2 \right|^2 \\
    \leq &2 
    \left( \int^{(i+1)h_1}_{ih_1} \left|\frac{\partial y_{p}(s_1,z_2)}{\partial s_1}\right| ds_1 \right)^2 \\ 
    & +2 \left( \int^{(j+1)h_2}_{jh_2} \left| \frac{\partial y_{p}(ih_1,s_2)}{\partial s_2} \right| ds_2\right)^2 \\
    \leq& 2h_1 \int_{ih_1}^{(i+1)h_1} \left| \frac{\partial y_{p} (s_1,z_2)}{\partial s_1} \right|^2 ds_1 \\
    &+ 2h_2 \int_{jh_2}^{(j+1)h_2} \left| \frac{\partial y_{p}(ih_1,s_2)}{\partial s_2} \right|^2 ds_2
\end{array}
\end{equation}
where the last inequality holds by using Cauchy-Schwarz inequality, i.e., $(\int_{a}^{b} \phi_1(x)\phi_2(x)dx)^2\leq \int_a^b (\phi_1(x))^2dx\int_a^b (\phi_2(x))^2dx$.

Let $Z_1=N_1h_1$, $Z_2=N_2h_2$, and we have
\[
\begin{array}{rl}
& \int^{Z_1}_0 \int^{Z_2}_0 \left|\Delta y_{p}(z_1,z_2)\right|^2 dz_2dz_1= \sum^{N_1-1}_{i=0}\sum^{N_2-1}_{j=0} \\
&  \int^{(i+1)h_1}_{ih_1}\int^{(j+1)h_2}_{jh_2} \left| y_{p}(z_1,z_2)-\hat{y}_{p}(i,j)\right|^2 dz_2dz_1 \\
\end{array}
\]
\[
\begin{array}{rl}
=&\sum^{N_1-1}_{i=0}\sum^{N_2-1}_{j=0} \int^{(i+1)h_1}_{ih_1}\int^{(j+1)h_2}_{jh_2} | y_{p}(z_1,z_2)-\hat{y}_{p}(i,z_2)\\
&+\hat{y}_{p}(i,z_2)-\hat{y}_{p}(i,j)|^2 dz_2dz_1,\\
\end{array}
\]
\[
\begin{array}{rl}
=&\sum^{N_1-1}_{i=0}\sum^{N_2-1}_{j=0} \int^{(i+1)h_1}_{ih_1}\int^{(j+1)h_2}_{jh_2} |
\int^{z_1}_{ih_1} \frac{\partial y_{p}(s_1,z_2)}{\partial s_1}ds_1\\
&+\int^{z_2}_{jh_2}\frac{\partial y_{p}(ih_1,s_2)}{\partial s_2} ds_2 |^2 dz_2dz_1.
\end{array}
\]

It follows from \eqref{eq:2-D caucgy scgward} that

\[
\begin{array}{rl}
 & \int^{Z_1}_0 \int^{Z_2}_0 \left|\Delta y_{p}(z_1,z_2)\right|^2 dz_2dz_1\\
 \end{array}
\]
\[
\begin{array}{rl}
\leq & 2 \sum^{N_1-1}_{i=0}\sum^{N_2-1}_{j=0} \int^{(i+1)h_1}_{ih_1}\int^{(j+1)h_2}_{jh_2}  h_1 \left\lgroup\right. \int^{(i+1)h_1}_{ih_1}  \\
&\left| \frac{\partial y_{p}(s_1,z_2)}{\partial s_1} \right|^2 ds_1 \left.\right\rgroup +h_2( \int^{(j+1)h_2}_{jh_2} \left| \frac{\partial y_{p}(ih_1,s_2)}{\partial s_2} \right|^2 ds_2 ) \\
& dz_2dz_1 \\
\end{array}
\]
\[
\begin{array}{rl}
\leq & 2 \sum^{N_1-1}_{i=0}\sum^{N_2-1}_{j=0} \left\lgroup\right. h_1^2 \int^{(j+1)h_2}_{jh_2}\int^{(i+1)h_1}_{ih_1} \left| \frac{\partial y_{p}(s_1,z_2)}{\partial s_1} \right|^2 \\
& ds_1dz_2 + h_2h_2^2 \int^{(j+1)h_2}_{jh_2} \left| \frac{\partial y_{p}(ih_1,s_2)}{\partial s_2} \right|^2 ds_2 \left.\right\rgroup.
\end{array}
\]

Next, it can be observed that
\[
\begin{array}{rl}
& 2h_1^2 \sum_{i=0}^{N_1-1} \sum_{j=0}^{N_2-1} \int^{(j+1)h_2}_{jh_2} \int^{(i+1)h_1}_{ih_1} \left| \frac{\partial y_{p}(s_1,z_2)}{\partial s_1} \right|^2 ds_1dz_2 \\
= & 2h_1^2 \int_{0}^{Z_2} \int_{0}^{Z_1} \left| \frac{\partial y_{p}(s_1,z_2)}{\partial s_1} \right|^2 ds_1dz_2\\
\overset{(a)}{\le}& 2h^{2}_1\alpha_1 ^{2} \int_{0}^{Z_1} \int_{0}^{Z_2} \left|u_{p}(z_1,z_2)\right|^2dz_2dz_1
\end{array}
\]
and
\[
\begin{array}{rl}
    &2h_1h_2^2 \sum^{N_1-1}_{i=0} \sum_{j=0}^{N_2-1} \int^{(j+1)h_2}_{jh_2} \left| \frac{\partial y_{p}(ih_1,s_2)}{\partial s_2} \right|^2 ds_2 \\
    =&2h_1h_2^2 \sum^{N_1-1}_{i=0} \int^{Z_2}_{0} \left| \frac{\partial y_{p}(ih_1,s_2)}{\partial s_2} \right|^2 ds_2 \\
     \overset{(b)}{\leq} & 2h_2^2 \sum_{i=0}^{N_1-1} h_1 \left[  \alpha_2^2 \int^{Z_2}_{0} \left| u_{p}(ih_1,z_2) \right|^2 dz_2 \right]\\
     \overset{(c)}{\leq} & 2h_2^2 \alpha_2^2 \int_{0}^{Z_1} \int_{0}^{Z_2} \left|u_p(z_1,z_2)\right|^2dz_2dz_1
\end{array}
\]
where (a)-(b) follow from Assumption \ref{asmp: good approximation} and (c) follows from the fact that the input $u_{p}(z_1,z_2)$ is set to be a piecewise signal because of the ZOH.
Hence it can be obtained that
\begin{align*}
&\int^{Z_1}_0 \int^{Z_2}_0 \left|\Delta y_{p}(z_1,z_2)\right|^2 dz_2dz_1 \\ 
 \leq & 2(\alpha_1^2h_1^2+\alpha_2^2h_2^2) \int^{Z_1}_0 \int^{Z_2}_0 \left|u_{p}(z_1,z_2)\right|^2 dz_2dz_1.
\end{align*}

\end{IEEEproof}

\subsection{Proof for Theorem \ref{thm:2-D passivity index}}
\begin{IEEEproof}
Let $Z_{1}=N_{1}h_{1}$ and $Z_{2}=N_{2}h_{2}$ where $N_{1},N_{2}\in\mathbb{N}$. 

First, let us denote the $QSR$ supply rates for systems $G_{p}$ and its sampled model $\hat{G}_{h}$ as 
\[
\begin{array}{c}
\omega_{p}(u_{p}(z_{1},z_{2}),y_{p}(z_{1},z_{2}))\triangleq y_{p}^{T}(z_{1},z_{2})Q_{p}y_{p}(z_{1},z_{2})\\
+2y_{p}^{T}(z_{1},z_{2})S_{p}u_{p}(z_{1},z_{2})+u_{p}^{T}(z_{1},z_{2})R_{p}u_{p}(z_{1},z_{2}),
\end{array}
\]
 and 
\[
\begin{array}{c}
\hat{\omega}_{p}(\hat{u}_{p}(i,j),\hat{y}_{p}(i,j))\triangleq\hat{y}_{p}^{T}(i,j)\hat{Q}_{p}\hat{y}_{p}(i,j)+\\
2\hat{y}_{p}^{T}(i,j)\hat{S}_{p}\hat{u}_{p}(i,j)+\hat{u}_{p}^{T}(i,j)\hat{R}_{p}\hat{u}_{p}(i,j),
\end{array}
\]
respectively.

Since $G_{p}$ is $(Q_{p},S_{p},R_{p})$-dissipative, there exists a well-defined non-negative storage function $V=V_{h}+V_{v}$ such that for all $N_{1},N_{2}\in\mathbb{N}$, 
\[
\int_{0}^{Z_{1}}\int_{0}^{Z_{2}}\omega_{p}\left(u_{p}(z_{1},z_{2}),y_{p}(z_{1},z_{2})\right)dz_{2}dz_{1}\ge\eta,
\]
where $\eta\triangleq\int_{0}^{Z_{1}}[V_{v}(x_{v}(z_{1},Z_{2}))-V_{v}(x_{v}(z_{1},0))]dz_{1}+\int_{0}^{Z_{2}}[V_{h}(x_{h}(Z_{1},z_{2}))-V_{h}(x_{h}(0,z_{2}))]dz_{2}.$ 

Next, from \eqref{eq:ZOH}-\eqref{eq:2-D delta y}, it can be obtained that
\[
\begin{array}{rl}
 & h_{1}h_{2}\sum_{i=0}^{N_{1}-1}\sum_{j=0}^{N_{2}-1}\hat{\omega}_{p}\left(\hat{u}_{p}(i,j),\hat{y}_{p}(i,j)\right)\\
= & h_{1}h_{2}\sum_{i=0}^{N_{1}-1}\sum_{j=0}^{N_{2}-1}[\hat{y}_{p}^{T}(i,j)\hat{Q}_{p}\hat{y}_{p}(i,j)+\\
 & 2\hat{y}_{p}^{T}(i,j)\hat{S}_{p}\hat{u}_{p}(i,j)+\hat{u}_{p}^{T}(i,j)\hat{R}_{p}\hat{u}_{p}(i,j)]\\
= & \int_{0}^{Z_{1}}\int_{0}^{Z_{2}}[y_{p}^{T}(z_{1},z_{2})\hat{Q}_{p}y_{p}(z_{1},z_{2})+\Delta y_{p}^{T}(z_{1},z_{2})\\
 & \hat{Q}_{p}\Delta y_{p}(z_{1},z_{2})-2\Delta y_{p}^{T}(z_{1},z_{2})\hat{Q}_{p}y_{p}(z_{1},z_{2})+\\
 & 2y_{p}^{T}(z_{1},z_{2})\hat{S}_{p}u_{p}(z_{1},z_{2})-2\Delta y_{p}^{T}(z_{1},z_{2})\hat{S}_{p}u_{p}(z_{1},z_{2})\\
 & +u_{p}^{T}(z_{1},z_{2})\hat{R}_{p}u_{p}(z_{1},z_{2})]dz_{2}dz_{1},
\end{array}
\]
which follows that 
\[
\begin{array}{l}
\;\;h_{1}h_{2}\sum_{i=0}^{N_{1}-1}\sum_{j=0}^{N_{2}-1}\hat{\omega}_{p}(\hat{u}_{p}(i,j),\hat{y}_{p}(i,j))-\\
 \;\; \int_{0}^{Z_{1}}\int_{0}^{Z_{2}}\omega_{p}(u_{p}(z_{1},z_{2}),y_{p}(z_{1},z_{2}))dz_{2}dz_{1}\\
=  \int_{0}^{Z_{1}}\int_{0}^{Z_{2}}[y_{p}^{T}(z_{1},z_{2})(\hat{Q}_{p}-Q_{p})y_{p}(z_{1},z_{2})+\Delta y_{p}^{T}(z_{1},z_{2})\\
\;\; \hat{Q}_{p}\Delta y_{p}(z_{1},z_{2})-2\Delta y_{p}^{T}(z_{1},z_{2})\hat{Q}_{p}y_{p}(z_{1},z_{2})+\\
 \;\; 2y_{p}^{T}(z_{1},z_{2})(\hat{S}_{p}-S_{p})u_{p}(z_{1},z_{2})-2\Delta y_{p}^{T}(z_{1},z_{2})\hat{S}_{p}u_{p}(z_{1},z_{2})\\
 \;\; +u_{p}^{T}(z_{1},z_{2})(\hat{R}_{p}-R_{p})u_{p}(z_{1},z_{2})]dz_{2}dz_{1}.
\end{array}
\]

Let us observe that 
\[
\begin{array}{rl}
 & -2\Delta y_{p}^{T}(z_{1},z_{2})\hat{Q}_{p}y_{p}(z_{1},z_{2})\\
\ge & -\xi_{1}\left|\hat{Q}_{p}y_{p}(z_{1},z_{2})\right|^{2}-\frac{1}{\xi_{1}}\left|\Delta y_{p}(z_{1},z_{2})\right|^{2},\vspace{2mm}\\

 & 2y_{p}^{T}(z_{1},z_{2})(\hat{S}_{p}-S_{p})u_{p}(z_{1},z_{2})\\
\ge & -\xi_{2}\left|(\hat{S}_{p}-S_{p})^{T}y_{p}(z_{1},z_{2})\right|^{2}-\frac{1}{\xi_{2}}\left|u_{p}(z_{1},z_{2})\right|^{2},\vspace{2mm}\\

 & -2\Delta y_{p}^{T}(z_{1},z_{2})\hat{S}_{p}u_{p}(z_{1},z_{2})\\
\ge & -\xi_{3}\left|\hat{S}_{p}u_{p}(z_{1},z_{2})\right|^{2}-\frac{1}{\xi_{3}}\left|\Delta y_{p}(z_{1},z_{2})\right|^{2},\\
\end{array}
\]
and
\[
\begin{array}{l}
\Delta y_{p}^{T}(z_{1},z_{2})\hat{Q}_{p}\Delta y_{p}(z_{1},z_{2}) \ge  \left|\underline{\lambda}(\hat{Q}_{p})\right|\left|\Delta y_{p}(z_{1},z_{2})\right|^{2},
\end{array}
\]
where $\xi_{1},\xi_{2},\xi_{3}$ can be any positive constants. Thus, it can be derived that 
\[
\begin{array}{rl}
 & h_{1}h_{2}\sum_{i=0}^{N_{1}-1}\sum_{j=0}^{N_{2}-1}\hat{\omega}_{p}\left(\hat{u}_{p}(i,j),\hat{y}_{p}(i,j)\right)-\\
 & \int_{0}^{Z_{1}}\int_{0}^{Z_{2}}\omega_{p}\left(u_{p}(z_{1},z_{2}),y_{p}(z_{1},z_{2})\right)dz_{2}dz_{1}\\
\ge & \int_{0}^{Z_{1}}\int_{0}^{Z_{2}}\left[(\underline{\lambda}(\hat{R}_{p}-R_{p})-\xi_{3}\bar{\sigma}_{max}^{2}(\hat{S}_{p})-\frac{1}{\xi_{2}})|u_{p}(z_{1},z_{2})|^{2}\right.\\
 & +(\underline{\lambda}(\hat{Q}_{p}-Q_{p})-\xi_{1}\bar{\sigma}_{max}^{2}(\hat{Q}_{p})-\xi_{2}\bar{\sigma}_{max}^{2}(\hat{S}_{p}-S_{p}))\times\\
 & |y_{p}(z_{1},z_{2})|^{2}\left.-(|\underline{\lambda}(\hat{Q}_{p})|+\frac{1}{\xi_{1}}+\frac{1}{\xi_{3}})|\Delta y_{p}(z_{1},z_{2})|^{2}\right]dz_{2}dz_{1}\\
\ge & \int_{0}^{Z_{1}}\int_{0}^{Z_{2}}\left[\left(\underline{\lambda}(\hat{R}_{p}-R_{p})-\xi_{3}\bar{\sigma}_{max}^{2}(\hat{S}_{p})-\frac{1}{\xi_{2}}-\right.\right.\\
 & \left.2(\alpha_{1}^2h_{1}^2+\alpha_{2}^2h_{2}^2)(|\underline{\lambda}(\hat{Q}_{p})|+\frac{1}{\xi_{1}}+\frac{1}{\xi_{3}})\right)|u_{p}(z_{1},z_{2})|^{2}+\\
 & \left.\left(\underline{\lambda}(\hat{Q}_{p}-Q_{p})-\xi_{1}\bar{\sigma}_{max}^{2}(\hat{Q}_{p})-\xi_{2}\bar{\sigma}_{max}^{2}(\hat{S}_{p}-S_{p})\right)\right.\\
 &\left. |y_{p}(z_{1},z_{2})|^{2}\right]dz_{2}dz_{1},
\end{array}
\]
where the last inequality follows from Lemma \ref{lem:2-D delta_y^2<ru^2}. If the condition in \eqref{eq:2-D passivity index} is satisfied, then the right hand side of the above inequality is non-negative,
which follows that 
\[
\sum_{i=0}^{N_{1}-1}\sum_{j=0}^{N_{2}-1}\hat{\omega}_{p}(\hat{u}_{p}(i,j),\hat{y}_{p}(i,j))\ge\frac{1}{h_{1}h_{2}}\eta.
\]
Since the sampled system $\hat{G}_{p}$ captures the same dynamical behavior as system $G_{p}$, by letting $\frac{1}{h_{1}h_{2}}\eta$ be the addiction of the storage function of $\hat{G}_{p}$, it is proved that $\hat{G}_{p}$ is $(\hat{Q}_{p},\hat{S}_{p},\hat{R}_{p})$-dissipative.
\end{IEEEproof}

\subsection{Proof for Theorem \ref{thm:linear passivity index}}
\begin{IEEEproof}
 Based on the solution of state trajectories
 $\left(x_{h}(z_1,z_2),x_{v}(z_1,z_2)\right)$, the dynamic of the sampled discrete linear 2-D system can be obtained as
\[
\begin{array}{l}
\label{eq:2-DDT system}
\left[
 \begin{matrix}
 x_{h}((i+1),j)\\
 x_{v}(i,(j+1))
 \end{matrix}
 \right]=\vspace{1mm}\\
 \left[
 \begin{matrix}
 e^{A_{11}h_{1}} & (e^{A_{11}h_{1}}-I)A_{11}^{-1}A_{12}\\
 (e^{A_{22}h_{2}}-I)A_{22}^{-1}A_{21} & e^{A_{22}h_{2}}
 \end{matrix}
 \right]\left[
 \begin{matrix}
 x_{h}(i,j)\\
 x_{v}(i,j)
 \end{matrix}
 \right]\\
 + \left[
 \begin{matrix}
 (e^{A_{11}h_{1}}-I)A_{11}^{-1}B_{1}\\
 (e^{A_{22}h_{1}}-I)A_{22}^{-1}B_{2}
 \end{matrix}
 \right]u(i,j),
 \\

 y(i,j)=\left[
 \begin{matrix}
 C_{1} & C_{2}
 \end{matrix}
 \right]\left[
 \begin{matrix}
 x_{h}(i,j)\\
 x_{v}(i,j)
 \end{matrix}
 \right]+Du(i,j),

\end{array}
\]
 and the sampled boundary condition becomes $x_{h}(0,j)$ and $x_{v}(i,0)$, $i,j\in\mathbb{N}$.  It can be observed that the input-output mapping of the discrete 2-D system above coincides with \eqref{eq:2-DCTsystem} when $z_1=ih_{1},z_2=jh_{2}, \forall i,j\in\mathbb{N}$. 

Let $Z_{1}=N_{1}h_{1}$ and $Z_{2}=N_{2}h_{2}$ where $N_{1},N_{2}\in\mathbb{N}$. 
Suppose there exist $P_{h},P_{v}>0$ such that the condition in \eqref{eq:$QSR$ LMI}
holds. Then, it can be implied that 
\[
\begin{array}{l}
x_{h}^{T}(i+1,j)P_{h}x_{h}(i+1,j)-x_{h}^{T}(i,j)P_{h}x_{h}(i,j)+\\
x_{v}^{T}(i,j+1)P_{v}x_{v}(i,j+1)-x_{v}^{T}(i,j)P_{v}x_{v}(i,j)\le
\\
 y_{p}^{T}(i,j)\hat{Q}_{p}y_{p}(i,j)+2y_{p}^{T}(i,j)\hat{S}_{p}u_{p}(i,j)+u_{p}^{T}(i,j)\hat{R}_{p}u_{p}(i,j)
\end{array}
\]
for all $x_{h}(i,j)\in\mathbb{R}^{n_{h}},x_{v}(i,j)\in\mathbb{R}^{n_{v}}$ and
all $u_{p}(i,j)\in\mathbb{R}^{m}$. By operating summation on both sides
of the above inequality, one has
\[
{\displaystyle \begin{array}{rl}
 & \begin{array}{l}
{\displaystyle {\displaystyle \sum_{j=0}^{N_{2}-1}}}\left[x_{h}^{T}(N_{1},j)P_{h}x_{h}(N_{1},j)-x_{h}^{T}(0,j)P_{h}x_{h}(0,j)\right]\\
{\displaystyle +\sum_{i=0}^{N_{1}-1}}\left[x_{v}^{T}(i,N_{2})P_{v}x_{v}(i,N_{2})-x_{v}^{T}(i,0)P_{v}x_{v}(i.0)\right]
\end{array}\\
\le & {\displaystyle \sum_{i=0}^{N_{1}-1}{\displaystyle {\displaystyle \sum_{j=0}^{N_{2}-1}}}}\left[y_{p}^{T}(i,j)\hat{Q}_{p}y_{p}(i,j)+2y_{p}^{T}(i,j)\hat{S}_{p}u_{p}(i,j)\right.\\
 & \left.+u_{p}^{T}(i,j)\hat{R}_{p}u_{p}(i,j)\right].
\end{array}}
\]

Therefore, by exploiting the storage function $V(x_{h}(i,j),x_{v}(i,j))=V_{h}(x_{h}(i,j))+V_{v}(x_{v}(i,j))$
where $V_{h}(x_{h}(i,j))=x_{h}^{T}(i,j)P_{h}x_{h}(i,j)$ and $V_{v}(x_{v}(i,j))=x_{v}^{T}(i,j)P_{v}x_{v}(i,j)$,
we can conclude, based on Definition \ref{def:2-D DT}, that
the sampled system $\hat{G}_{p}$ is $(\hat{Q}_{p},\hat{S}_{p},\hat{R}_{p})$-dissipative.
\end{IEEEproof}
\subsection{Proof for Theorem \ref{thm:2-D passivity index with quantization}}

\begin{IEEEproof}
For quantizer $\mathcal{Q}_p$ with  input $\hat{y}_p$, we have
\begin{align}
\label{eq:2-D quantizer Delta y < delta*y}
    \left|\Delta y_p(i,j)\right| \leq \delta_p \left|\hat{y}_p(i,j)\right|,
\end{align}
where the quantization error is defined by $\Delta y_p(i,j)\triangleq\mathcal{Q}_p(\hat{y}_p(i,j))-\hat{y}_p(i,j)$. Similarly, for quantizer $\mathcal{Q}_c$ with input $y_c$ with quantization error denoted as $\Delta y_c(i,j)\triangleq\mathcal{Q}_c(y_c(i,j))-y_c(i,j)$, we have 
\begin{align}
    \label{eq:2-D quantizer Delta y < delta*y 2}
        \left|\Delta y_c(i,j)\right| \leq \delta_c \left|y_c(i,j)\right|.
\end{align}
Let $(x_{h}^{p},x_{v}^{p})$ and $(x_{h}^{c},x_{v}^{c})$ be the states of system $\hat{G}_{p}$ and $G_{c}$, respectively. Since the system $\hat{G}_{p}$ is IF-OFP$(\rho_{p},\nu_{p})$, there exists a non-negative storage function $V^{p}=V_{h}^{p}+V_{v}^{p}$ such that for any $N_1,N_2 \in \mathbb{N}$,
\begin{equation}\label{eq:omega_p,eta_p}
{\displaystyle \sum_{i=0}^{N_{1}-1}{\displaystyle {\displaystyle \sum_{j=0}^{N_{2}-1}}}}\omega_{p}(\hat{u}_{p}(i,j),\hat{y}_{p}(i,j))\ge\eta_{p},
\end{equation}
where
\begin{equation*}
\begin{array}{l}
\omega_{p}\left(\hat{u}_{p}(i,j),\hat{y}_{p}(i,j)\right)=-\rho_{p}|\hat{y}_{p}(i,j)|^{2}+\\
\;\;\hat{y}_{p}^{T}(i,j)\hat{u}_{p}(i,j)-\nu_{p}|\hat{u}_{p}(i,j)|^{2},\\
\end{array}
\end{equation*}
\begin{equation*}
\begin{array}{l}
\eta_{p}= { \sum_{i=0}^{N_{1}-1}}\left[V_{v}^{p}\left(x_{v}^{p}(i,N_{2})\right)-V_{v}^{p}\left(x_{v}^{p}(i,0)\right)\right]+\\
\;\;{ \sum_{j=0}^{N_{2}-1}}\left[V_{h}^{p}\left(x_{h}^{p}(N_{1},j)\right)-V_{h}^{p}\left(x_{h}^{p}(0,j)\right)\right].
\end{array}
\end{equation*}
 
Similarly, since $G_{c}$ is IF-OFP$(\rho_{c},\nu_{c})$,
there exists a non-negative storage function $V^{c}=V_{h}^{c}+V_{v}^{c}$ such
that for any $N_1,N_2 \in \mathbb{N}$,
\begin{equation}\label{eq:omega_c,eta_c}
{\displaystyle \sum_{i=0}^{N_{1}-1}{\displaystyle {\displaystyle \sum_{j=0}^{N_{2}-1}}}}\omega_{c}\left(u_{c}(i,j),y_{c}(i,j)\right)\ge\eta_{c},
\end{equation}
where 
\begin{equation*}
\begin{array}{l}
\omega_{c}\left(u_{c}(i,j),y_{c}(i,j)\right)= -\rho_{c}|y_{c}(i,j)|^{2}+\\
  \;\; y_{c}^{T}(i,j)u_{c}(i,j)-\nu_{c}|u_{c}(i,j)|^{2},\\
\end{array}
\end{equation*}
\begin{equation*}
\begin{array}{l}
\eta_{c}= { \sum_{i=0}^{N_{1}-1}}\left[V_{v}^{c}\left(x_{v}^{c}(i,N_{2})\right)-V_{v}^{c}\left(x_{v}^{c}(i,0)\right)\right]+\\
\;\;{ \sum_{j=0}^{N_{2}-1}}\left[V_{h}^{c}\left(x_{h}^{c}(N_{1},j)\right)-V_{h}^{c}\left(x_{h}^{c}(0,j)\right)\right].
\end{array}
\end{equation*}

For the rest of this proof, we will omit the ordered pair notation $(i,j)$ for notational simplicity. Since $\hat{u}_{p}=r_{p}-\mathcal{Q}_{c}(y_{c})$, $u_{c}=r_{c}+\mathcal{Q}_{p}(\hat{y}_{p})$,
$\mathcal{Q}_{p}(\hat{y}_{p})=\Delta y_{p}+\hat{y}_{p}$ and $\mathcal{Q}_{c}(y_{c})=\Delta y_{c}+y_{c}$,
we have 
\[
\begin{array}{rl}
 & \omega_{p}(\hat{u}_{p},\hat{y}_{p})+\omega_{c}(u_{c},y_{c})\\
= & -\rho_{p}\hat{y}_{p}^{T}\hat{y}_{p}-\nu_{p}\left(r_{p}^{T}r_{p}-2\mathcal{Q}_{c}^{T}(y_{c})r_{p}+\mathcal{Q}_{c}^{T}(y_{c})\mathcal{Q}_{c}(y_{c})\right)\\
 & +\hat{y}_{p}^{T}r_{p}-\hat{y}_{p}^{T}\mathcal{Q}_{c}(y_{c})-\rho_{c}y_{c}^{T}y_{c}-\nu_{c}\left(r_{c}^{T}r_{c}+2\mathcal{Q}_{p}^{T}(\hat{y}_{p})r_{c}\right.\\
 & \left. +\mathcal{Q}_{p}^{T}(\hat{y}_{p})\mathcal{Q}_{p}(\hat{y}_{p})\right)+y_{c}^{T}r_{c}+y_{c}^{T}\mathcal{Q}_{p}(\hat{y}_{p})\\
= & \begin{bmatrix}\hat{y}_{p}^T & y_{c}^T \end{bmatrix}\begin{bmatrix}-(\rho_{p}+\nu_{c})I & 0\\
0 & -(\rho_{c}+\nu_{p})I
\end{bmatrix}\begin{bmatrix}\hat{y}_{p}\\
y_{c}
\end{bmatrix}\\
& +2\begin{bmatrix}\hat{y}_{p}^T & y_{c}^T \end{bmatrix}\begin{bmatrix}\frac{1}{2}I & -\nu_{c}I \\ \nu_{p}I & \frac{1}{2}I \end{bmatrix}\begin{bmatrix}r_{p} \\ r_{c} \end{bmatrix}\\
& +\begin{bmatrix}r_{p}^T & r_{c}^T \end{bmatrix}\begin{bmatrix}-\nu_{p}I & 0 \\ 0 & -\nu_{c}I \end{bmatrix}\begin{bmatrix}r_{p} \\ r_{c} \end{bmatrix}\\
& +2\nu_{p}\Delta y_{c}^{T}r_{p}-\nu_{p}|\Delta y_{c}|^{2}-2\nu_{p}\Delta y_{c}^{T}y_{c}-\hat{y}_{p}^{T}\Delta y_{c}\\
& -2\nu_{c}\Delta y_{p}^{T}r_{c}-\nu_{c}|\Delta y_{p}|^{2}-2\nu_{c}\Delta y_{p}^{T}\hat{y}_{p}+y_{c}^{T}\Delta y_{p}.
\end{array}
\]

It follows from \eqref{eq:2-D quantizer Delta y < delta*y}- \eqref{eq:2-D quantizer Delta y < delta*y 2} that 
\[
\left\{ \begin{array}{rcl}
-\hat{y}_{p}^{T}\Delta y_{c} & \le & \frac{1}{2\beta_{1}}|\hat{y}_{p}|^{2}+\frac{\beta_{1}}{2}\delta_{c}^{2}|y_{c}|^{2}\\
y_{c}^{T}\Delta y_{p} & \le & \frac{1}{2\beta_{2}}|y_{c}|^{2}+\frac{\beta_{2}}{2}\delta_{p}^{2}|\hat{y}_{p}|^{2}\\
-2\nu_{p}\Delta y_{c}^{T}y_{c} & \le & 2\delta_{c}|\nu_{p}||y_{c}|^{2}\\
-2\nu_{c}\Delta y_{p}^{T}\hat{y}_{p} & \le & 2\delta_{p}|\nu_{c}||\hat{y}_{p}|^{2}\\
2\nu_{p}\Delta y_{c}^{T}r_{p} & \le & \delta_{c}^{2}|y_{c}|^{2}+|\nu_{p}|^{2}|r_{p}|^{2}\\
-2\nu_{c}\Delta y_{p}^{T}r_{c} & \le & \delta_{p}^{2}|\hat{y}_{p}|^{2}+|\nu_{c}|^{2}|r_{c}|^{2}
\end{array}\right.
\]
where $\beta_{1},\beta_{2}$ could be any positive constants. 

Thus, it can be obtained that
\[
\begin{array}{rl}
& \omega_{p}(\hat{u}_{p},\hat{y}_{p})+\omega_{c}(u_{c},y_{c})\\ 
\le & \begin{bmatrix}\hat{y}_{p}\\y_{c} \end{bmatrix}^{T} \begin{bmatrix}q_{1}I & 0 \\ 0 & q_{2}I \end{bmatrix}\begin{bmatrix}\hat{y}_{p}\\
y_{c} \end{bmatrix}+2\begin{bmatrix}\hat{y}_{p} \\ y_{c} \end{bmatrix}^{T} \begin{bmatrix} \frac{1}{2}I & -\nu_{c}I \\ \nu_{p}I & \frac{1}{2}I \end{bmatrix}\begin{bmatrix}r_{p} \\ r_{c} \end{bmatrix}\\
& +\begin{bmatrix}r_{p}\\r_{c} \end{bmatrix}^{T}\begin{bmatrix}r_{1}I & 0\\ 0 & r_{2}I \end{bmatrix}\begin{bmatrix}r_{p}\\ r_{c}\end{bmatrix},
\end{array}
\]
where
\begin{equation}\label{eq:q1q2}
\left\{ \begin{array}{c}
q_{1}=-(\rho_{p}+\nu_{c})+(\delta_{p}^{2}+2\delta_{p})|\nu_{c}|+(1+\frac{\beta_{2}}{2})\delta_{p}^{2}+\frac{1}{2\beta_{1}}\\
q_{2}=-(\rho_{c}+\nu_{p})+(\delta_{c}^{2}+2\delta_{c})|\nu_{p}|+(1+\frac{\beta_{1}}{2})\delta_{c}^{2}+\frac{1}{2\beta_{2}}
\end{array}\right.,
\end{equation}
\begin{equation}\label{eq:r1r2}
\left\{ \begin{array}{c}
r_{1}=-\nu_{p}+|\nu_{p}|^{2}\\
r_{2}=-\nu_{c}+|\nu_{c}|^{2}
\end{array}\right..
\end{equation}

Let us denote 
\[
Q'=\begin{bmatrix}q_{1}I & 0\\
0 & q_{2}I
\end{bmatrix},S'=\begin{bmatrix}\frac{1}{2}I & -\nu_{c}I\\
\nu_{p}I & \frac{1}{2}I
\end{bmatrix}, R'=\begin{bmatrix}r_{1}I & 0\\
0 & r_{2}I
\end{bmatrix}.
\]

 It can be observed that 
\[
\begin{array}{l}
\eta_{p}+\eta_{c}
\le  {\displaystyle \sum_{i=0}^{N_{1}-1}}{\displaystyle {\displaystyle \sum_{j=0}^{N_{2}-1}}}\omega_{p}(\hat{u}_{p},\hat{y}_{p})+\omega_{c}(u_{c},y_{c})\le \\
 {\displaystyle \sum_{i=0}^{N_{1}-1}}{\displaystyle {\displaystyle \sum_{j=0}^{N_{2}-1}}}
 \begin{bmatrix}\hat{y}_{p}\\y_{c}\end{bmatrix}^{T}Q'\begin{bmatrix}\hat{y}_{p}\\y_{c}\end{bmatrix} 
 + 2\begin{bmatrix}\hat{y}_{p}\\y_{c}\end{bmatrix}^{T}S'\begin{bmatrix}r_{p}\\r_{c}\end{bmatrix} 
 + \begin{bmatrix}r_{p}\\r_{c}\end{bmatrix}^{T}R'\begin{bmatrix}r_{p}\\r_{c}\end{bmatrix}.
\end{array}
\]

By letting the storage function of the closed-loop system be $V=V_{h}+V_{v}$
where $V_{h}=V_{h}^{p}+V_{h}^{c}$ and $V_{v}=V_{v}^{p}+V_{v}^{c}$,
it can be concluded that the interconnected system is $(Q',S',R')$-dissipative.

Therefore, based on Lemma \ref{lem:2-D Q<0 implies l2 stable}, we can conclude that the closed-loop system under quantization is $\mathcal{L}_2$-stable if $q_1,q_2<0$ where $q_1$ and $q_2$ are defined in \eqref{eq:q1q2}. 
\end{IEEEproof}

\subsection{Proof for Theorem \ref{thm: 2-D triggering}}
\begin{IEEEproof}
Since the systems $\hat{G}_{p}$and $G_{c}$ are IF-OFP$(\rho_{p},\nu_{p})$ and IF-OFP$(\rho_{c},\nu_{c})$, respectively, there exist non-negative storage functions $V^{p}=V_{h}^{p}+V_{v}^{p}$ and $V^{c}=V_{h}^{c}+V_{v}^{c}$ satisfying \eqref{eq:omega_p,eta_p}-\eqref{eq:omega_c,eta_c} correspondingly.

Based on $\hat{u}_{p}(i,j)=r_{p}(i,j)-\mathcal{Q}_{c}\left(y_{c}(i,j)\right)$
and $u_{c}(i,j)=r_{c}(i,j)+\mathcal{Q}_{p}\left(\hat{y}_{p}(i,j_{k})\right)$,
it follows that 
\begin{align*}
& \quad  \omega_{p}\left(\hat{u}_{p}(i,j),\hat{y}_{p}(i,j)\right)+\omega_{c}\left(u_{c}(i,j),y_{c}(i,j)\right)\\
&=  -\rho_{p}\hat{y}_{p}^{T}(i,j)\hat{y}_{p}(i,j)-\nu_{p}\left[ r_{p}^{T}(i,j)r_{p}(i,j)\right.
\\ 
& \quad \left.
-2\mathcal{Q}_{c}^{T}(y_{c}(i,j))r_{p}(i,j)+\mathcal{Q}_{c}^{T}(y_{c}(i,j))\mathcal{Q}_{c}(y_{c}(i,j))\right]\\
& \quad +\hat{y}_{p}^{T}(i,j)r_{p}(i,j)-\hat{y}_{p}^{T}(i,j)\mathcal{Q}_{c}(y_{c}(i,j))-\rho_{c}y_{c}^{T}(i,j)y_{c}(i,j)\\
& \quad -\nu_{c}\left[r_{c}^{T}(i,j)r_{c}(i,j)+2\mathcal{Q}_{p}^{T}(\hat{y}_{p}(i,j_{k}))r_{c}(i,j)\right.\\
& \quad \left.+\mathcal{Q}_{p}^{T}(\hat{y}_{p}(i,j_{k}))\mathcal{Q}_{p}(\hat{y}_{p}(i,j_{k}))\right]+y_{c}^{T}(i,j)r_{c}(i,j)  \\
&\quad +y_{c}^{T}(i,j)\mathcal{Q}_{p}(\hat{y}_{p}(i,j_{k})).
\end{align*}
Following the proof in Theorem \ref{thm:2-D passivity index with quantization} and $\mathcal{Q}_{p}(\hat{y}_{p}(i,j_{k}))=\hat{y}_{p}(i,j)+\Delta y_{p}(i,j)-e(i,j)$ where $\Delta y_p(i,j)\triangleq\mathcal{Q}_p(\hat{y}_p(i,j))-\hat{y}_p(i,j)$,
it can be derived that 

\[
\begin{array}{rl}
&\;\;\omega_{p}(\hat{u}_{p}(i,j),\hat{y}_{p}(i,j))+\omega_{c}(u_{c}(i,j),y_{c}(i,j))\\
\le & \begin{bmatrix}\hat{y}_{p}(i,j)^T & y_{c}(i,j)^T \end{bmatrix} \begin{bmatrix}q_{1}I & 0 \\ 0 & q_{2}I \end{bmatrix} \begin{bmatrix}\hat{y}_{p}(i,j) \\ y_{c}(i,j) \end{bmatrix} \\
& +2\begin{bmatrix}\hat{y}_{p}(i,j)^T & y_{c}(i,j)^T \end{bmatrix} \begin{bmatrix}\frac{1}{2}I & -\nu_{c}I\\\nu_{p}I & \frac{1}{2}I \end{bmatrix} \begin{bmatrix}r_{p}(i,j)\\r_{c}(i,j)\end{bmatrix}\\
&+\begin{bmatrix}r_{p}(i,j)^T & r_{c}(i,j)^T \end{bmatrix}\begin{bmatrix}r_{1}I & 0\\0 & r_{2}I \end{bmatrix} \begin{bmatrix}r_{p}(i,j)\\r_{c}(i,j)\end{bmatrix}\\
& +2\nu_{c}e^{T}(i,j)r_{c}(i,j)-\nu_{c}|e(i,j)|^{2}-y_{c}^{T}(i,j)e(i,j)\\
& +2\nu_{c}\mathcal{Q}_{p}^{T}(\hat{y}_{p}(i,j))e(i,j),
\end{array}
\]
where $q_{1},q_{2},r_{1},r_{2}$ are defined in \eqref{eq:q1q2}-\eqref{eq:r1r2}. Since $2\nu_{c}e^{T}(i,j)r_{c}(i,j)\le \left|\nu_{c}\right|\left|e(i,j)\right|^{2}+\left|\nu_{c}\right|\left|r_{c}(i,j)\right|^{2}$,
we can further have 
\[
\begin{array}{l}
\omega_{p}(\hat{u}_{p}(i,j),\hat{y}_{p}(i,j))+\omega_{c}(u_{c}(i,j),y_{c}(i,j))\
\end{array}
\]
\[
\begin{array}{l}
\le  y^{T}(i,j)\bar{Q}y(i,j)+2y^{T}(i,j)\bar{S}r(i,j)+r^{T}(i,j)\bar{R}r(i,j)+\\
 (|\nu_{c}|-\nu_{c})|e(i,j)|^{2}-y_{c}^{T}(i,j)e(i,j)+2\nu_{c}\mathcal{Q}_{p}^{T}(\hat{y}_{p}(i,j))e(i,j),
\end{array}
\]
where $y\triangleq\begin{bmatrix}\hat{y}_{p}^{T} & y_{c}^{T}\end{bmatrix}^{T}$,
$r\triangleq\begin{bmatrix}r_{p}^{T} & r_{c}^{T}\end{bmatrix}^{T}$, $\bar{Q}=\begin{bmatrix}q_{1}I & 0\\
0 & q_{2}I
\end{bmatrix}$, $\bar{S}=\begin{bmatrix}\frac{1}{2}I & -\nu_{c}I\\\nu_{p}I & \frac{1}{2}I\end{bmatrix}$ and $\bar{R}=\begin{bmatrix}r_{1}I & 0\\0 & (r_{2}+|\nu_{c}|)I
\end{bmatrix}$.

Next, let $\theta_{1}$ and $\theta_{2}$ be any constant within $(0,1)$. Under the condition that $q_{1},q_{2}<0$, it can be verified that
\[
\begin{array}{l}
y^{T}(i,j)\bar{Q}y(i,j)+2y^{T}(i,j)\bar{S}r(i,j)+r^{T}(i,j)\bar{R}r(i,j)+\\
(|\nu_{c}| -\nu_{c})|e(i,j)|^{2}-y_{c}^{T}(i,j)e(i,j)+2\nu_{c}\mathcal{Q}_{p}^{T}(\hat{y}_{p}(i,j))e(i,j)\\
\end{array}
\]
\[
\begin{array}{l}
=  y^{T}(i,j)\begin{bmatrix}\begin{array}{r}
(1-\theta_{1})q_{1}I\end{array} & 0\\0 & (1-\theta_{2})q_{2}I \end{bmatrix}y(i,j)\\
\quad +2y^{T}(i,j)\bar{S}r(i,j) +r^{T}(i,j)\bar{R}r(i,j)\\
\quad -\left|\sqrt{-\theta_{2}q_{2}}y_{c}(i,j)+\frac{1}{2\sqrt{-\theta_{2}q_{2}}}e(i,j)\right|^{2} +\theta_{1}q_{1}|\hat{y}_{p}(i,j)|^{2}\\
\quad +\left(|\nu_{c}|-\nu_{c}-\frac{1}{4\theta_{2}q_{2}}\right)|e(i,j)|^{2}+2\nu_{c}\mathcal{Q}_{p}^{T}(\hat{y}_{p}(i,j))e(i,j).
\end{array}
\]

Denote $\epsilon^{2}=|\nu_{c}|-\nu_{c}-\frac{1}{4\theta_{2}q_{2}}$
and $Q=\begin{bmatrix}\begin{array}{r}
(1-\theta_{1})q_{1}I\end{array} & 0\\
0 & (1-\theta_{2})q_{2}I
\end{bmatrix}$, then one has 
\[
\begin{array}{l}
y^{T}(i,j)\bar{Q}y(i,j)+2y^{T}(i,j)\bar{S}r(i,j)+r^{T}(i,j)\bar{R}r(i,j)\\
+(|\nu_{c}|-\nu_{c})|e(i,j)|^{2}-y_{c}^{T}(i,j)e(i,j)\\
+2\nu_{c}\mathcal{Q}_{p}^{T}(\hat{y}_{p}(i,j))e(i,j)\\
\end{array}
\]
\[
\begin{array}{l}
\le  y^{T}(i,j)Qy(i,j)+2y^{T}(i,j)\bar{S}r(i,j)+r^{T}(i,j)\bar{R}r(i,j)\\
\quad +\epsilon^{2}|e(i,j)|^{2}+2\nu_{c}\mathcal{Q}_{p}^{T}(\hat{y}_{p}(i,j))e(i,j)+\theta_{1}q_{1}|\hat{y}_{p}(i,j)|^{2}\\
\end{array}
\]
\[
\begin{array}{l}
=  y^{T}(i,j)Qy(i,j)+2y^{T}(i,j)\bar{S}r(i,j)+r^{T}(i,j)\bar{R}r(i,j)\\
\quad -\frac{\nu_{c}^{2}}{\epsilon^{2}}|\mathcal{Q}_{p}(\hat{y}_{p}(i,j))|^{2}+\theta_{1}q_{1}|\hat{y}_{p}(i,j)|^{2}\\
\quad +\left|\epsilon e(i,j)+\frac{\nu_{c}}{\epsilon}\mathcal{Q}_{p}(\hat{y}_{p}(i,j))\right|^{2}
\end{array}
\]
\[
\begin{array}{l}
\le  y^{T}(i,j)Qy(i,j)+2y^{T}(i,j)\bar{S}r(i,j)+r^{T}(i,j)\bar{R}r(i,j)\\
\quad -\left(\frac{\nu_{c}^{2}}{\epsilon^{2}}-\frac{\theta_{1}q_{1}}{(1+\delta_{p})^{2}}\right)|\mathcal{Q}_{p}(\hat{y}_{p}(i,j))|^{2}\\
\quad +\left|\epsilon e(i,j)+\frac{\nu_{c}}{\epsilon}\mathcal{Q}_{p}(\hat{y}_{p}(i,j))\right|^{2},
\end{array}
\]
where the last inequality is obtained from the observation that $|\mathcal{Q}_{p}(\hat{y}_{p})|^{2}\le(1+\delta_{p})^{2}|\hat{y}_{p}|^{2}$
based on \eqref{eq:bound_quantizer}. Hence, it can be concluded that
for any $N_{2}\in\mathbb{N}$, 

\[
\begin{array}{l}
 \;\; {\displaystyle \sum_{i=0}^{N_{1}-1}\sum_{j=0}^{N_{2}-1}}\omega_{p}(\hat{u}_{p}(i,j),\hat{y}_{p}(i,j))+\omega_{c}(u_{c}(i,j),y_{c}(i,j))\\
\le  {\displaystyle \sum_{i=0}^{N_{1}-1}\sum_{j=0}^{N_{2}-1}}y^{T}(i,j)Qy(i,j)+2y^{T}(i,j)\bar{S}r(i,j)+
\\
\;\; r^{T}(i,j)\bar{R}r(i,j)+ \left|\epsilon e(i,j)+\frac{\nu_{c}}{\epsilon}\mathcal{Q}_{p}(\hat{y}_{p}(i,j))\right|^{2}-
\\
\;\;\left(\frac{\nu_{c}^{2}}{\epsilon^{2}}-\frac{\theta_{1}q_{1}}{(1+\delta_{p})^{2}}\right)|\mathcal{Q}_{p}(\hat{y}_{p}(i,j))|^{2}.
\end{array}
\]
Given the triggering condition \eqref{eq:triggering condition}, it
can be implied that for any $j\in\{j_{k}+1,\ldots,j_{k+1}\}$,
\[
\;\;{\displaystyle \begin{array}{l}
{\displaystyle \sum_{i=0}^{N_{1}-1}}\left|\epsilon e(i,j)+\frac{\nu_{c}}{\epsilon}\mathcal{Q}_{p}(\hat{y}_{p}(i,j))\right|^{2}\\
\le\left(\frac{\nu_{c}^{2}}{\epsilon^{2}}-\frac{\theta_{1}q_{1}}{(1+\delta_{p})^{2}}\right){\displaystyle \sum_{i=0}^{N_{1}-1}}|\mathcal{Q}_{p}(\hat{y}_{p}(i,j))|^{2}
\end{array}}
\]
which indicates that 
\[
\begin{array}{l}
 \;\;\; {\displaystyle \sum_{i=0}^{N_{1}-1}\sum_{j=0}^{N_{2}-1}}\omega_{p}(\hat{u}_{p}(i,j),\hat{y}_{p}(i,j))+\omega_{c}(u_{c}(i,j),y_{c}(i,j))\\
\le {\displaystyle \sum_{i=0}^{N_{1}-1}\sum_{j=0}^{N_{2}-1}}y^{T}(i,j)Qy(i,j)+2y^{T}(i,j)\bar{S}r(i,j)
\\
\;\;\;\; +r^{T}(i,j)\bar{R}r(i,j).
\end{array}
\]

By letting the storage function of the closed-loop system be $V=V_{h}+V_{v}$
where $V_{h}=V_{h}^{p}+V_{h}^{c}$ and $V_{v}=V_{v}^{p}+V_{v}^{c}$,
it can be obtained that 
\[
\begin{array}{l}
\;\;{ \sum_{i=0}^{N_{1}-1}}[V_{v}^{p}(x_{v}^{p}(i,N_{2}))+V_{v}^{c}(x_{v}^{c}(i,N_{2}))-V_{v}^{p}(x_{v}^{p}(i,0))\\
\;\; -V_{v}^{c}(x_{v}^{c}(i,0))]+{ \sum_{j=0}^{N_{2}-1}}[V_{h}^{p}(x_{h}^{p}(N_{1},j))+ V_{h}^{c}(x_{h}^{c}(N_{1},j))\\
\;\;-V_{h}^{p}(x_{h}^{p}(0,j))-V_{h}^{c}(x_{h}^{c}(0,j))]\\
\le{ \sum_{i=0}^{N_{1}-1}\sum_{j=0}^{N_{2}-1}}y^{T}(i,j)Qy(i,j)+2y^{T}(i,j)\bar{S}r(i,j)\\
\;\;\;+r^{T}(i,j)\bar{R}r(i,j)\\
\le{ \sum_{i=0}^{N_{1}-1}\sum_{j=0}^{N_{2}-1}}a|y(i,j)|^{2}+2b|y(i,j)||r(i,j)|+c|r(i,j)|^{2}\\
\le{ \sum_{i=0}^{N_{1}-1}\sum_{j=0}^{N_{2}-1}}\frac{1}{2a}(a|y(i,j)|+2b|r(i,j)|)^{2}+\frac{a}{2}|y(i,j)|^{2}\\
\;\; \;+\left(c-\frac{2b^{2}}{a}\right)|r(i,j)|^{2}\\
\le{ \sum_{i=0}^{N_{1}-1}\sum_{j=0}^{N_{2}-1}}\frac{a}{2}|y(i,j)|^{2}+\left(c-\frac{2b^{2}}{a}\right)|r(i,j)|^{2},
\end{array}
\]
where $a\triangleq\text{max}\{(1-\theta_{1})q_{1},(1-\theta_{2})q_{2}\}<0$,
$b\triangleq\sigma_{max}(\bar{S})\geq 0$, and $c\triangleq\sigma_{max}(\bar{R}) \geq 0$.
Thus, we have 
\[
{\displaystyle \sum_{i=0}^{N_{1}-1}\sum_{j=0}^{N_{2}-1}|y(i,j)|^{2}\le{\displaystyle \sum_{i=0}^{N_{1}-1}\sum_{j=0}^{N_{2}-1}}\frac{4b^{2}-2ac}{a^{2}}|r(i,j)|^{2}-\frac{2\kappa}{a}},
\]
where $\kappa={ \sum_{i=0}^{N_{1}-1}[V_{v}^{p}(x_{v}^{p}(i,0))+V_{v}^{c}(x_{v}^{c}(i,0))]}+{ \sum_{j=0}^{N_{2}-1}}[V_{h}^{p}(x_{h}^{p}(0,j))+V_{h}^{c}(x_{h}^{c}(0,j))]$.
Hence, according to Definition \ref{def: bounded l2 -2d}, the interconnected
system in Figure \ref{fig:2DNCS} has a finite $\mathcal{L}_{2}$-gain,
which completes the proof.
\end{IEEEproof}

%
\bibliographystyle{IEEEtran}
\bibliography{References}

\begin{IEEEbiography}[{\includegraphics[width=1in,height=1.25in,clip,keepaspectratio]{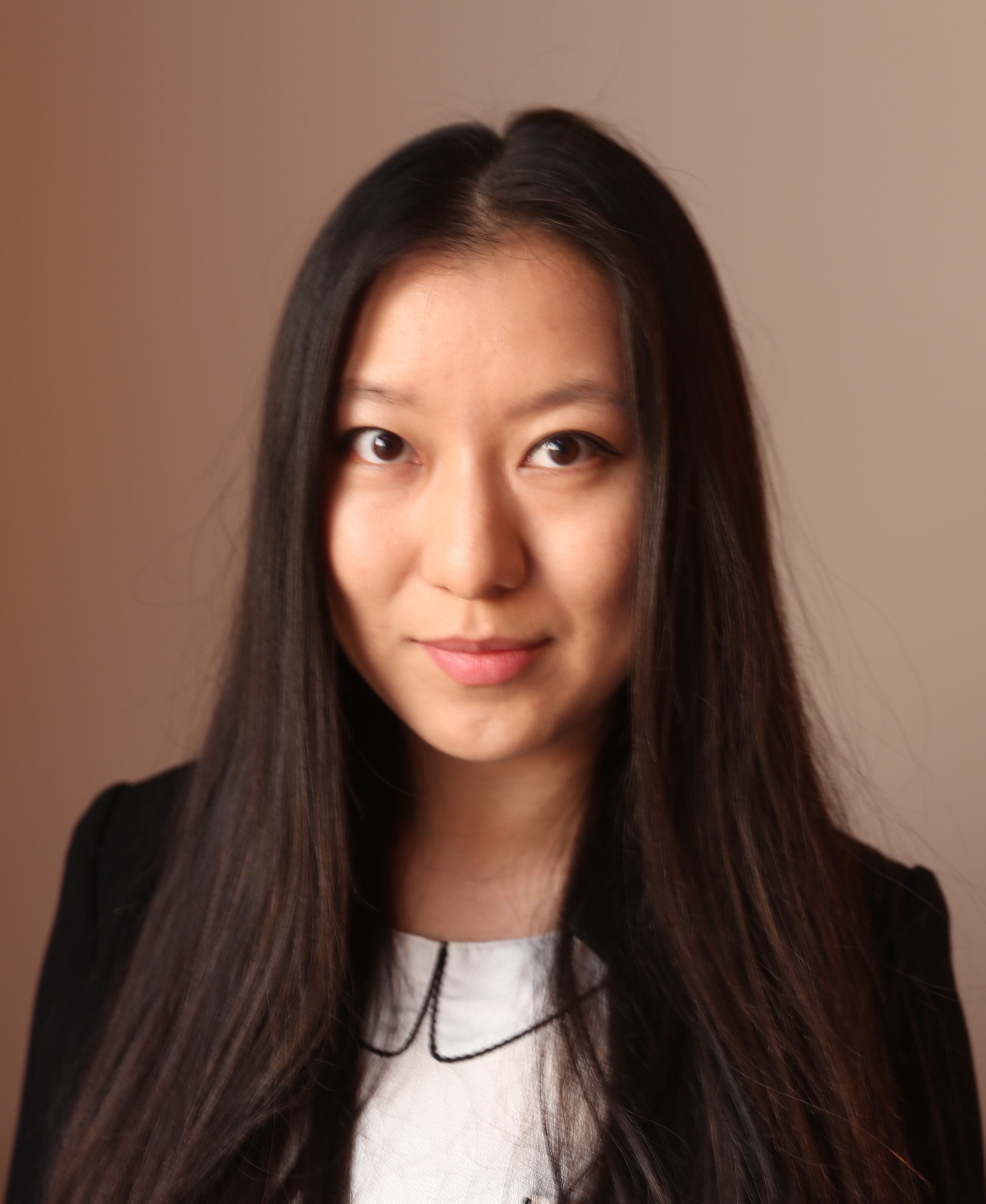}}]{Yang Yan}
is currently a Ph.D. candidate in the Department of Electrical Engineering at the University of Notre Dame. She received the B.S. degree in Measurement, Control and Instrument in 2012 and the M.S. degree in Control Science and Engineering in 2014, both from Shandong University, China. Her research addresses problems of control and automation and examines ways to design engineering system with numerical methods. Her research interests include Cyber-Physical Systems, optimization algorithms and networked control systems.
\end{IEEEbiography}

\begin{IEEEbiography}[{\includegraphics[width=1in,height=1.25in,clip,keepaspectratio]{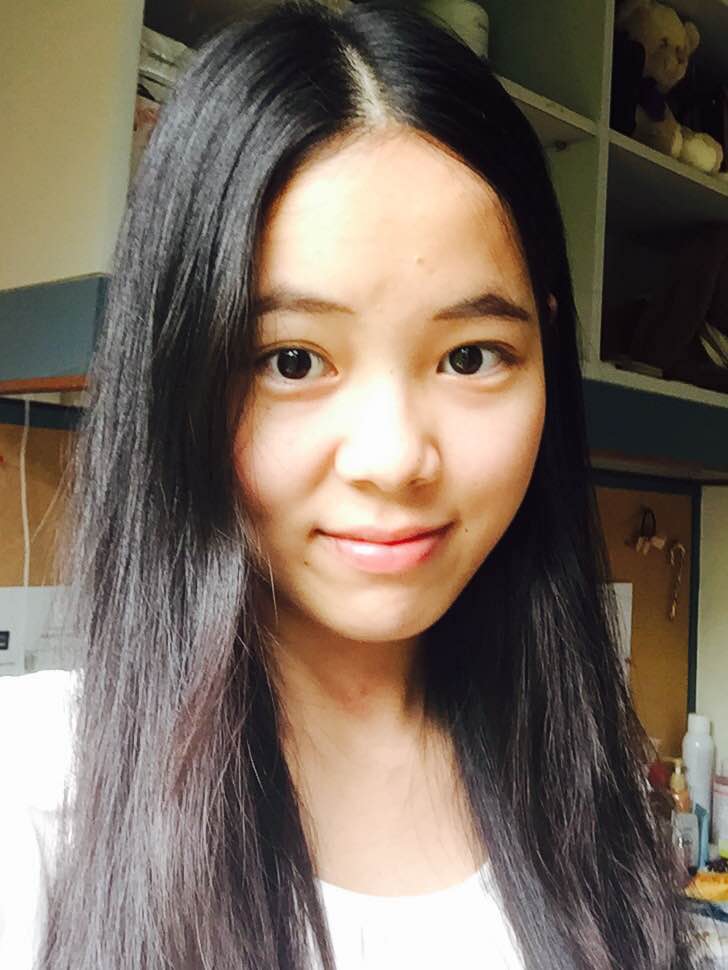}}]{Lanlan Su} received the B.E. degree in electrical engineering from Zhejiang University, China, in 2014, and the Ph.D degree in  electrical and electronic engineering from The University of Hong Kong, in 2018.  She is currently a postdoctoral research associate in University of Notre Dame. She also is an awardee of the Hong Kong Ph.D. Fellowship Scheme established by the Research Grants Council of Hong Kong. Dr. Su$'$s research interests include robust cotrol, dissipativity, networked control system and distributed optimization. 
\end{IEEEbiography}

\begin{IEEEbiography}[{\includegraphics[width=1in,height=1.25in,clip,keepaspectratio]{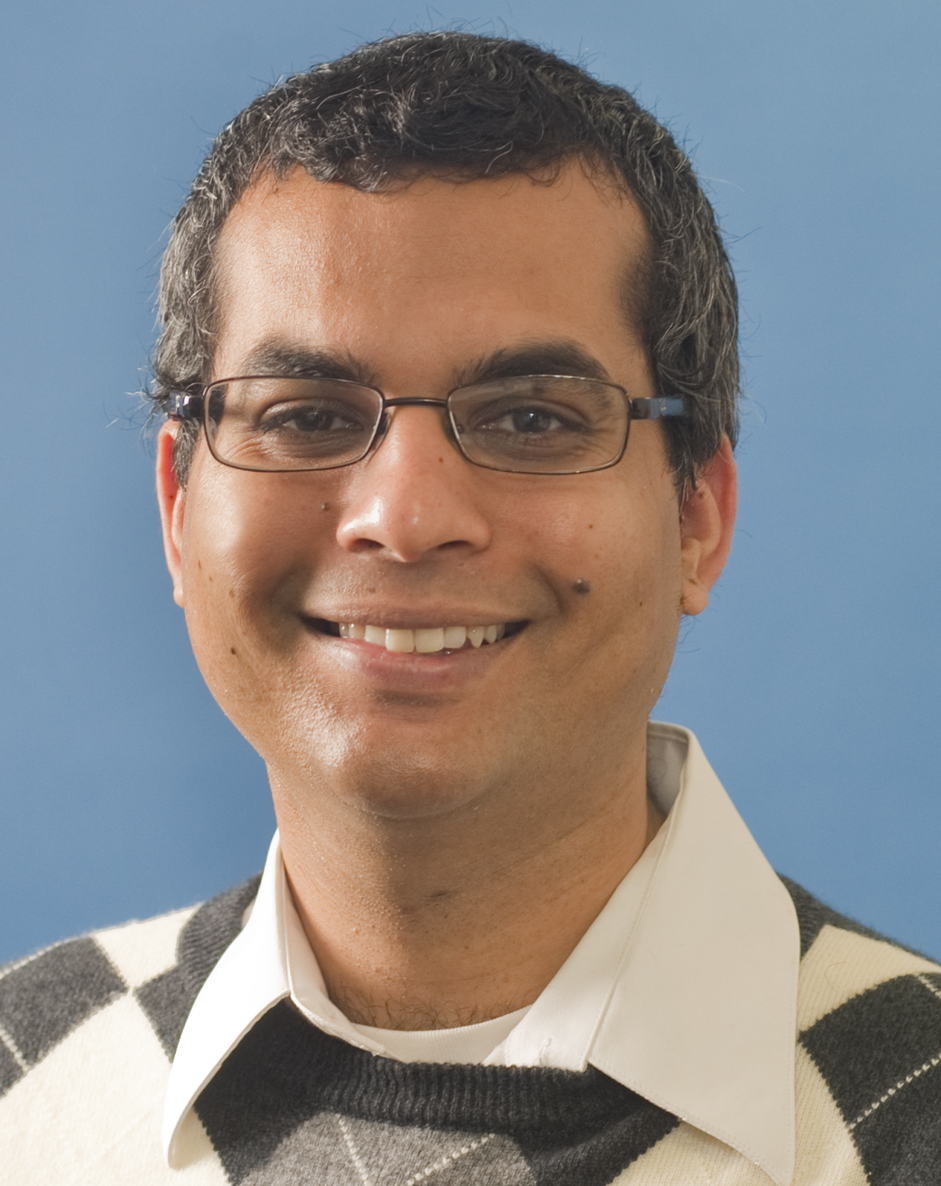}}]{Vijay Gupta}
is a Professor in the Department of Electrical Engineering at the University of Notre Dame, having joined the faculty in January 2008. He received his B. Tech degree at Indian Institute of Technology, Delhi, and his M.S. and Ph.D. at California Institute of Technology, all in Electrical Engineering. Prior to joining Notre Dame, he also served as a research associate in the Institute for Systems Research at the University of Maryland, College Park. He received the 2013 Donald P. Eckman Award from the American Automatic Control Council and a 2009 National Science Foundation (NSF) CAREER Award. His research and teaching interests are broadly at the interface of communication, control, distributed computation, and human decision making.
\end{IEEEbiography}

\begin{IEEEbiography}[{\includegraphics[width=1in,height=1.25in,clip,keepaspectratio]{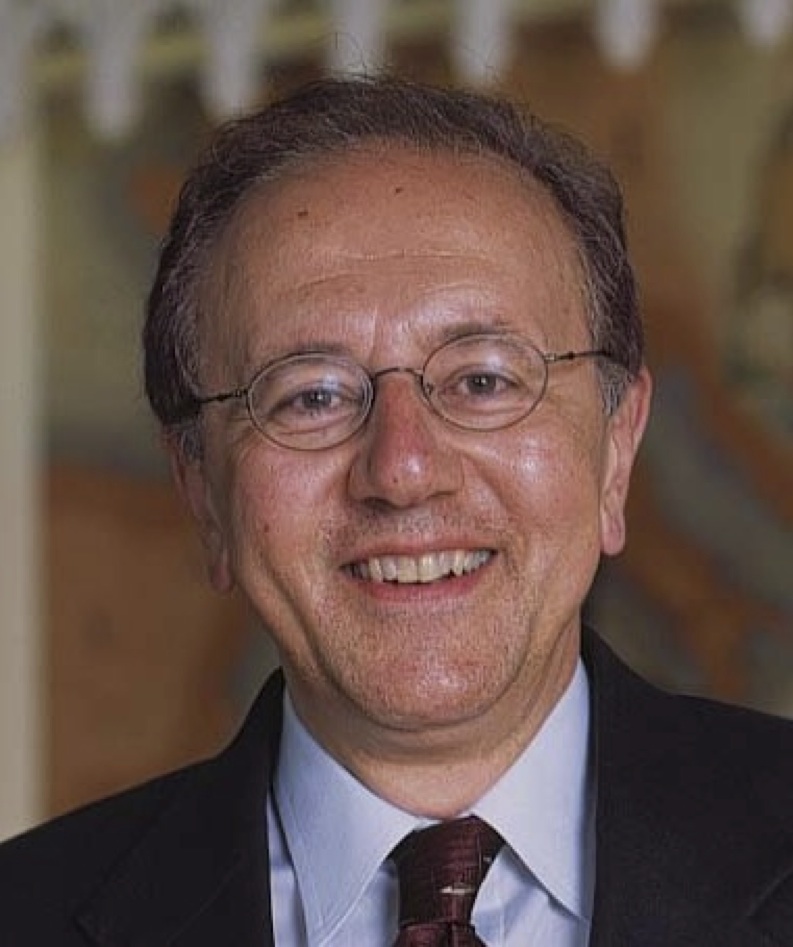}}]{Panos Antsaklis}
is the H.C. \& E.A. Brosey Professor of Electrical Engineering at the University of Notre Dame. He is graduate of the National Technical University of Athens, Greece, and holds MS and PhD degrees from Brown University. His research addresses problems of control and automation and examines ways to design control systems that will exhibit high degree of autonomy. His current research focuses on Cyber-Physical Systems and the interdisciplinary research area of control, computing and communication networks, and on hybrid and discrete event dynamical systems. He is IEEE, IFAC and AAAS Fellow, President of the Mediterranean Control Association, the 2006 recipient of the Engineering Alumni Medal of Brown University and holds an Honorary Doctorate from the University of Lorraine in France. He served as the President of the IEEE Control Systems Society in 1997 and was the Editor-in-Chief of the IEEE Transactions on Automatic Control for 8 years, 2010-2017.
\end{IEEEbiography}

\end{document}